\pdfoutput=1

\documentclass[12pt]{article}
\usepackage{amsmath,epsf,amssymb,latexsym,cite,shadow,eucal,theorem,enumerate}
\usepackage[matrix,arrow,frame,import,curve,color]{xy}
\usepackage{tikz}
\usetikzlibrary{calc}
\usetikzlibrary{arrows}

\usepackage{graphicx}

\newtheorem{theorem}{Theorem}
\newtheorem{prop}{Proposition}

\newtheorem{conjecture}{Conjecture}

{\theorembodyfont{\rmfamily}}

\newif\iffigs\figstrue

\DeclareFontFamily{U}{rsf}{}
\DeclareFontShape{U}{rsf}{m}{n}{
  <5> <6> rsfs5 <7> <8> <9> rsfs7 <10-> rsfs10}{}
\DeclareMathAlphabet\Scr{U}{rsf}{m}{n}

\def\O{\Scr{O}}
\def\C{{\mathbb C}}
\def\P{{\mathbb P}}

\def\Q{{\mathbb Q}}

\def\Z{{\mathbb Z}}

\def\sHom{\operatorname{\Scr{H}\!\!\textit{om}}}

\def\Ext{\operatorname{Ext}}

\def\End{\operatorname{End}}

\def\Spec{\operatorname{Spec}}

\def\SO{\operatorname{SO}}
\def\Sl{\operatorname{SL}}
\def\Gl{\operatorname{GL}}

\def\SU{\operatorname{SU}}
\def\GU{\operatorname{U{}}}

\def\diag{\operatorname{diag}}

\def\ch{\operatorname{\mathit{ch}}}
\def\td{\operatorname{\mathit{td}}}

\def\nut{\widetilde{\nu}}

\def\gammab{\overline{\gamma}}

\def\CY{Calabi--Yau}

\def\cR{{\Scr R}}
\def\cM{{\Scr M}}

\def\cA{{\Scr A}}

\def\cK{{\Scr K}}

\def\cT{{\Scr T}}

\def\cE{{\Scr E}}

\def\cR{{\Scr R}}
\def\cW{{\Scr W}}

\def\ff#1#2{{\textstyle\frac{#1}{#2}}}

\def\poso#1{#1\save="x"!LD+<0pt,-0.5mm>;
  "x"!RD+<0pt,-0.5mm>**\dir{.}\restore}

\def\eqn#1#2{\begin{equation}#2
  \ifx{#1}{}\else\label{#1}\fi\end{equation}}

\begin{document}

\begin{titlepage}
\begin{flushright}
October 2011
\end{flushright}
\vspace{.5cm}
\begin{center}
\baselineskip=16pt
{\fontfamily{ptm}\selectfont\bfseries\huge
A McKay-Like Correspondence\\
for (0,2)-Deformations\\[20mm]}
{\bf\large  Paul S.~Aspinwall
 } \\[7mm]

{\small

Center for Geometry and Theoretical Physics, 
  Box 90318 \\ Duke University, 
 Durham, NC 27708-0318 \\ \vspace{6pt}

 }

\end{center}

\begin{center}
{\bf Abstract}
\end{center}
We present a local computation of deformations of the tangent bundle
for a resolved orbifold singularity $\C^d/G$. These correspond to
$(0,2)$-deformations of $(2,2)$-theories. A McKay-like correspondence
is found predicting the dimension of the space of first-order
deformations from simple calculations involving the group.  This is
confirmed in two dimensions using the Kronheimer--Nakajima quiver
construction. In higher dimensions such a computation is subject to
nontrivial worldsheet instanton corrections and some examples are given where
this happens. However, we conjecture that the special crepant resolution
given by the $G$-Hilbert scheme is never subject to such corrections,
and show this is true in an infinite number of cases. Amusingly, for
three-dimensional examples where $G$ is abelian, the moduli space is
associated to a quiver given by the toric fan of the blow-up. It
is shown that an orbifold of the form $\C^3/\Z_7$ has a nontrivial
superpotential and thus an obstructed moduli space.


\end{titlepage}

\vfil\break


\section{Introduction}    \label{s:intro}

Perhaps the first result in string theory to excite interest from
geometers arose from orbifolds. In \cite{DHVW:} it was shown that one
could straight-forwardly analyze string theory on $\C^d/G$, where $G$
is some finite subgroup of $\Sl(d,\C)$, despite the fact that this
orbifold is a singular space. It was shown that certain string states
are in one-to-one correspondence with conjugacy classes of $G$. If one
were to consider string theory on a smooth space $X$, such string
states would correspond to even-dimensional homology classes. Thus, if
$\C^d/G$ is deformed into some smooth space $X$, and there is reason
to expect the string spectrum is unchanged by such a resolution, one
predicts a relationship between the even-dimensional cohomology of $X$
and the conjugacy classes of $G$.

Such a statement is a consequence of the McKay correspondence in two
dimensions \cite{McKay:}. This has been
extended to dimension three in papers such as
\cite{MR1404917,BKM:MisM}.

It is interesting to note, however, that the original orbifold paper
\cite{DHVW:} also contained another numerical prediction for orbifolds
which has received a good deal less attention. This concerns the
deformation of $N=(2,2)$ theories to $(0,2)$ theories. This is phrased
in the language of heterotic string as follows. One may
compactify\footnote{In a loose sense. In this paper $X$ is not
  compact!} an $E_8\times E_8$ heterotic string on a \CY\ $n$-fold $X$
together with a principal $E_8\times E_8$ bundle $E\to X$. The
easiest way to do this is to use the ``standard embedding'' which was
very much in vogue in the early days of string theory. In this case,
one uses the tangent bundle $T$ for $E$ by embedding $\SU(d)$ into
$E_8\times E_8$.

Such a model has an underlying $N=(2,2)$ superconformal field
theory. There are deformations of this theory which preserve this
supersymmetry. If $X$ is smooth such deformations can be interpreted
as deforming the complex structure and complexified K\"ahler form of
$X$. Such marginal deformations are always truly marginal thanks to
the extended supersymmetry \cite{Dixon:}. This is exactly the same
statement as the unobstructedness of the moduli space as in
\cite{Tian:def,Tod:WP}.

More interestingly, as far as this paper is concerned, one may deform
to a conformal field theory with only $(0,2)$ supersymmetry. This
corresponds to a deformation of the compactification of the heterotic
string by deforming the tangent bundle to another bundle $E$. In the
context of the analysis of this paper, $E$ will always be
a holomorphic bundle with structure group $\SU(d)$.

From a phenomenological point of view, it is clearly absurd these days
to use the standard embedding. However, for a better
understanding of $(0,2)$ theories, it makes sense to begin
with the much better-understood $(2,2)$-theories and venture into the
world of $(0,2)$-theories through deformations. In this way, the
standard embedding becomes a very important idea.

String theory would appear to imply that there is a connection
between counting certain massless states in an orbifold conformal field theory
and the number of deformations of the tangent bundle on a crepant
resolution of the orbifold. We will dub this the ``(0,2)-McKay
Correspondence''. 

The unobstructedness theorems of $(2,2)$-deformations are no longer
valid for $(0,2)$-deformations when one deals with spaces of more than two
complex dimensions. In particular, three possibilities are of
interest:
\begin{enumerate}
\item The deformations of $E$ itself may be geometrically obstructed.
\item The mutual deformations of $X$ and the bundle $E$ may be
  obstructed.
\item There may be worldsheet instanton effects ruling out certain
  $E\to X$
  as true vacua.
\end{enumerate}
These possibilities all correspond to a nontrivial superpotential for
the scalar fields associated to the deformations of the theory. In
particular, quadratic terms in the superpotential affect masses and
thus change the number of massless states. The analysis of
(0,2)-deformations is therefore a good deal more interesting than
(2,2)-deformations. Of course, a pessimist would say that the
worldsheet instanton effects, which are included in the conformal
field theory but not in the geometric nonlinear $\sigma$-model, should
completely ruin such a (0,2)-McKay correspondence in dimension
$\geq3$. In this paper we shall show that a particular class of
orbifold resolution, namely the $G$-Hilbert scheme, appears to be
immune from such effects, at least in many cases.

After reviewing the counting of orbifold states in section
\ref{s:count} we will prove that the (0,2)-McKay correspondence works
perfectly in dimension two. This should come as no surprise to a
string theorist, but is an interesting mathematical result.

For the remainder of the paper we consider dimension three. In section
\ref{s:dim3} we develop machinery to compute the number of
deformations of the tangent sheaf on a resolved abelian orbifold. The
moduli space of the tangent sheaf is described in terms of a quiver
which, in a satisfying coincidence, is given in terms of the toric
diagram of the resolution itself. In section \ref{s:egs} we give
various examples and this motivates our conjecture that only a
particular resolution, the $G$-Hilbert scheme, need satisfy the
(0,2)-McKay correspondence.

In section \ref{s:superp} we show how a nontrivial cubic
superpotential can appear in a specific example, and finally
in section \ref{s:disc} we present concluding comments.


\section{Counting States in an Orbifold}  \label{s:count}

In this section we review how to count the number of massless states
in an orbifold theory on $\C^d/G$ that can be used to marginally
deform a $N=(2,2)$ superconformal symmetry to a theory with at least
$N=(0,2)$ supersymmetry. We call such states ``singlets'' from their
r\^ole in heterotic compactifications.

Let $x^i$ be the bosonic fields corresponding to the holomorphic
coordinates in the target space $\C^d$. Their (target space) complex
conjugates are $\bar x^i$. We have right moving fermions denoted
$\psi^i$ and $\bar\psi^i$ which are superpartners of $x^i$, $\bar x^i$
with respect to the right-moving $N=2$ worldsheet supersymmetry. We
have left-moving fermions $\gamma^i$,$\bar\gamma^i$. 

Let us consider fields in the twisted sector corresponding to an
element $g\in G$. First recall how we would do this computation if we were only
concerned with singlets which, when used as marginal operators,
preserved the full $N=(2,2)$ supersymmetry. These correspond to
(anti)chiral primary operators. Upon twisting the theory to a
topological field theory, these states are elements of BRST
cohomology. The spectrum of a twisted sector is accordingly given by
the Dolbeault or De Rahm (depending on the twisting) cohomology of the
fixed point set of $g$ \cite{Zas:}.

In our case we have a larger spectrum as we wish only to preserve
$N=(0,2)$ supersymmetry. One of the right-moving worldsheet
supersymmetry generators, $\bar Q$, can be used to play the r\^ole of
a BRST operator. We need only consider fields in representations of
the left-moving $N=2$ supersymmetry which are elements of cohomology
of $\bar Q$. This was the trick used in\cite{Kachru:1993pg} and
again in \cite{meMP:singlets}. Fixing on $\bar Q$-cohomology will
again localize information about the twisted sector to the fixed point
set. Our interest in this paper is in {\em isolated\/} fixed points
and thus the $\bar Q$-cohomology is trivial. In other words, we ignore
all right-moving oscillators. Thus the computation focuses purely on
the left-moving sector.

As an element of $\Sl(d,\C)$ let $g$ have
eigenvalues $\exp(2\pi i\nu_i)$, for $0\leq\nu_i<1$ and $i=1\ldots
d$. Following the notation of \cite{meMP:singlets} this implies we
have expansions for the left-movers:
\begin{equation}
\begin{split}
x^i(z) &= \sum_{u\in\Z-\nu_i} x_u^i z^{-u}\\
2\partial\bar x^i(z) &= \sum_{u\in\Z+\nu_i} \rho_u^i z^{-u-1}\\
\gamma^i(z) &=  \sum_{u\in\Z-\tilde\nu_i} \gamma_u^i z^{-u-\frac12}\\
\bar\gamma^i(z) &=  \sum_{u\in\Z+\tilde\nu_i} \bar\gamma_u^i z^{-u-\frac32}
\end{split} \label{eq:modes}
\end{equation}
where $\tilde\nu_i$ is defined as $\nu_i-\ff s2\pmod1$ such that
$-1<\tilde\nu_i\leq 0$; with the Ramond sector given by $s=0$ and the
Neveu-Schwarz sector given by $s=1$.

The energy of the twisted vacuum is \cite{DHVW:,meMP:singlets}
\begin{equation}
  E = \frac12\sum_{i=1}^d\Bigl(\nu_i(1-\nu_i) +
   \tilde\nu_i(1+\tilde\nu_i)\Bigr)+\frac d8 -1.
\end{equation}

The $N=(2,2)$ superconformal field theory has left and right $\GU(1)$
currents. For a heterotic compactification, the left-moving current is
part of the unbroken gauge symmetry whilst the right-moving current
corresponds to an R-charge. The vacuum charges are
\begin{equation}
\begin{split}
  q &= -\frac d2-\sum_i\tilde\nu_i\\
  \bar q &= -\frac d2+\sum_i\nu_i.
\end{split}
\end{equation}

From the vacuum we build states by applying operators from the
expansions (\ref{eq:modes}). The lowest modes we denote by
\begin{equation}
x_i \equiv x^i_{-\nu_i}, \quad \rho_i \equiv \rho^i_{\nu_i-1}, 
\quad \gamma_i \equiv \gamma^i_{-1-\nut_i}, \quad
\gammab_i \equiv \gammab^i_{\nut_i}.  \label{eq:lowest}
\end{equation}
Thus we have weights and charges given by table~\ref{table:charges}.
\begin{table}
\begin{center}
\begin{tabular}{|c|c|c|c|c|}
\hline
&$x_i$&$\rho_i$&$\gamma_i$&$\bar\gamma_i$\\
\hline
$E$&$\nu_i$&$1-\nu_i$&$1+\tilde\nu_i$&$-\tilde\nu_i$\\
$q$&0&0&$-1$&1\\
$\bar q$&0&0&0&0\\
\hline
\end{tabular}
\caption{Weights and charges of the fields}
\label{table:charges}
\end{center}
\end{table}

To preserve the $N=2$ right-moving supersymmetry when the singlet is
used as a marginal operator, when viewed as a right-moving chiral
primary field, it must have $\bar q=1$. We will work in the right-moving
Ramond sector. By spectral flow this means we are looking for 
states with
\begin{equation}
\bar q = 1-\frac d2. \label{eq:qbar}
\end{equation}

States in the orbifold must be invariant under the action of
$G$. Consider the action of $h\in G$ on a $g$-twisted state. The
result is a state in the $h^{-1}gh$-twisted sector. $G$-invariant
states therefore consist of orbits spanning the elements of each
conjugacy class of $G$. Each member of this orbit must be invariant under
elements $h\in G$ which commute with $g$.

Suppose $h$ commutes with $g$. We may assume the action of both $g$
and $h$ have been diagonalized on $x_i$. The action of $h$ on the
modes (\ref{eq:lowest}) is thus given. 

The $g$-twisted vacuum itself may transform nontrivially under $h$ as
analyzed in \cite{meMP:singlets}. We also need to restrict attention
to an odd fermion number for the GSO projection.  The condition
(\ref{eq:qbar}) is given by $\sum\nu_i=1$. In this case we may then
state the transformation rules as follows:
\begin{itemize}
\item If $\nu_i\leq\ff12$ for all $i$, then the vacuum is invariant
  and the NS-vacuum has odd fermion number.
\item If $\nu_j>\ff12$ (which can only be true for a single $j$) then
  the vacuum transforms like $\bar x_j$ and the NS-vacuum has even
  fermion number.
\end{itemize}

This gives a complete algorithm for computing the spectrum of massless
singlets on an orbifold. For each conjugacy class of $G$ we enumerate
states with $E=q=0$ and $\bar q$ given by (\ref{eq:qbar}) by applying
polynomials in the modes (\ref{eq:lowest}) to the vacuum which are
invariant under the centralizer and have odd fermion number.


\section{Dimension 2}  \label{s:dim2}

In this section we will prove the (0,2)-McKay correspondence in
dimension two by checking each possible case of $\C^2/G$ for $G$ a
finite subgroup of $\Sl(2,\C)$.  As is well-known, these finite groups
are in one-to-one correspondence with the Dynkin diagrams of $A_n$,
$D_n$ or $E_n$ and we label the groups accordingly. 

The moduli space of vector bundles is unobstructed in dimension two
\cite{Muk:K3}. In other words, there is too much supersymmetry for an
``interesting'' superpotential. This also implies that there can be no
worldsheet instanton corrections to the dimension of the moduli
space. It follows that string theory implies the (0,2)-McKay
correspondence must work.

In dimension two the moduli space has a quaternionic K\"ahler
structure. We do not use this fact but it motivates the counting of states
in terms of quaternionic degrees of freedom.

\subsection{The Orbifold Spectrum of States} \label{ss:orb2}

First we do the
orbifold computation.

\begin{theorem} \label{th:orb2}
For the orbifold $\C^2/G$, where $G$ corresponds to $A_n$, $D_n$ or $E_n$
as a subgroup of $\Sl(2,\C)$, we have $3n+1$ quaternionic singlets. 
\end{theorem}

Before we prove this theorem, let us note that we know from the
standard McKay correspondence that $n$ singlets correspond to blowing
up the singularity since the finite groups $A_n$, $D_n$ or $E_n$ each
have $n$ nontrivial conjugacy classes. Thus the theorem indicates that
we should expect there to be $2n+1$ singlets associated with deforming
the tangent bundle.  We now prove the theorem individually in each
case of $A_n$, $D_n$ or $E_n$.  From (\ref{eq:qbar}) we look for $\bar
q=0$ states.

Any nontrivial element $g\in G$ corresponds to $\nu_1+\nu_2=1$.  First
we consider the left-moving Ramond sector for which
$\tilde\nu_1=-\nu_2$ and $\tilde\nu_2=-\nu_1$. One easily computes
$E=q=\bar q=0$ but the vacuum has even fermion number. There are thus
no singlets intrinsic to this sector.\footnote{A heterotic string
  compactification on a complex surface with the standard embedding
  gives a six-dimensional theory with an $E_8\times E_7$ gauge
  symmetry. In this context we have left-moving fermions
  $\psi_\alpha$, $\alpha=1,\ldots,6$ from the $\SO(12)$ remnant of
  $E_8$ not involved in the conformal field theory. Odd products of
  these may be applied to the twisted vacuum to build a spinor
  $\mathbf{32}$ of $\SO(12)$. Other massless states build this up into
  a $\mathbf{56}$ of $E_7$ for each conjugacy class of $G$. This is
  again a statement of the conventional McKay correspondence.}

We can therefore focus purely on the NS-sector. We now consider each case.

\subsubsection{$\C^2/\Z_2$}

There is only one twisted sector. One computes for the vacuum
\begin{equation}
\begin{split}
\nu_i&=\ff12, \quad\tilde\nu_i=0\\
q&=-\sum(\tilde\nu_i+\ff12)=-1\\
\bar q&=\sum(\nu_i-\ff12)=0\\
E&=-\ff12.
\end{split}
\end{equation}

The modes have weights and charges

\renewcommand\arraystretch{1.3}
\begin{equation}
\begin{array}{|c|c|c|c|}
\hline
&E&q&\bar q\\
\hline
x_i&\ff12&0&0\\
\rho_i&\ff12&0&0\\
\gamma_i&1&-1&0\\
\bar\gamma_i&0&1&0\\
\hline
\end{array}
\end{equation}

This yields the following states with $E=q=\bar q=0$:
\begin{itemize}
\item $x_i\bar\gamma_j|g\rangle$ giving 4 singlets.
\item $\rho_i\bar\gamma_j|g\rangle$ giving 4 more singlets.
\end{itemize}
All of these are invariant under $g$ and have odd fermion number.
These states are complex. Therefore this counts as 4 quaternionic
singlets in total. This proves the theorem for $A_1$.

\subsubsection{$\C^2/\Z_q$}
Let $g\in \C^2/\Z_q$ have eigenvalues $\exp(2\pi i\nu_i)$ for
$=1,2$. If $\nu_1=\nu_2=\ff12$ we reduce to the $\Z_2$ case
above. Otherwise, by relabeling if necessary, we let $\nu_1=p/q$, where
$2p<q$.

The vacuum $|p,q\rangle$ then has
\begin{equation}
\begin{split}
q&=-\sum(\tilde\nu_i+\ff12)\\
\bar q&=\sum(\nu_i-\ff12)=0\\
E&=\ff pq-1
\end{split}
\end{equation}
while the excitation modes have
\renewcommand\arraystretch{1.3}
\begin{equation}
\begin{array}{|c|c|c|c|}
\hline
&E&q&\bar q\\
\hline
x_i&\ff pq, 1-\ff pq&0&0\\
\rho_i&1-\ff pq, \ff pq&0&0\\
\gamma_i&\ff pq+\ff 12,\ff12-\ff pq&-1&0\\
\bar\gamma_i&\ff12-\ff pq,\ff pq+\ff12&1&0\\
\hline
\end{array}
\end{equation}

This yields the following states
\begin{itemize}
\item $x_2|p,q\rangle, \rho_1|p,q\rangle, 
  \bar\gamma_1\gamma_2x_1|p,q\rangle,\bar\gamma_1\gamma_2\rho_2|p,q\rangle$.
  Always invariant.
\item $x_1^a\rho_2^b|p,q\rangle$ where $a+b=q-1$, but invariant only if $p=1$.
\end{itemize}

This counts as $4+q$ half-quaternionic singlets in total if $p=1$, and
4 otherwise.  Combining
all sectors in the $\Z_q$-orbifold we then have $3q-2$ complete
quaternionic singlets. That is, an $A_n$ singularity $\C^2/\Z_{n+1}$ contributes
$3n+1$ singlets.

\subsubsection{$\C^2/D_n$}

Let $g$ generate the group $\Z_{2(n-2)}$ and $h$ be such that
$g^{n-2}=h^2=-1$ and $ghg=h$.
\begin{itemize}
\item The $(-1)$-twisted sector needs to be invariant under all of
  $D_n$. Only a single quaternionic dimension remains.

\item The $g$-twisted sector (including the $g^{-1}$ conjugate) contributes
$n$ as above.

\item Each $g^2,\ldots,g^{n-3}$ sector contributes 2 as above.

\item There remain two conjugacy classes given by $h$ and $gh$. Each is like
the $\Z_4$-twist and each gives 4 quaternionic singlets.
\end{itemize}
This gives a total of $3n+1$.

\subsubsection{$\C^2/E_6$}

There are 7 conjugacy classes. One is the identity. Then:
\begin{itemize}
\item The $(-1)$-twisted sector contributes 1.
\item A single $\Z_4$ sector gives 4.
\item Two sectors are twisted by $\Z_6$. Each gives 5.
\item The squares of the above each give 2.
\end{itemize}
The total is 19.

\subsubsection{$\C^2/E_7$}

There are 8 conjugacy classes. One is the identity. Then:
\begin{itemize}
\item The $(-1)$-twisted sector contributes 1.
\item A single independent $\Z_4$ sector gives 4.
\item One sector is twisted by $\Z_8$. This gives 6. The square of
  this sector and cube each give 2 more.
\item One sector is twisted by $\Z_6$. This gives 5. The square of
  this gives 2.
\end{itemize}
The total is 22.

\subsubsection{$\C^2/E_8$}

There are 9 conjugacy classes. One is the identity. Then:
\begin{itemize}
\item The $(-1)$-twisted sector contributes 1.
\item A single independent $\Z_4$ sector gives 4.
\item One sector is twisted by $\Z_{10}$. This gives 7. The square of
  this sector and cube and 4th power each give 2 more.
\item One sector is twisted by $\Z_6$. This gives 5. The square of
  this gives 2.
\end{itemize}
The total is 25. 

\noindent This concludes the proof of theorem \ref{th:orb2}. $\Box$

\subsection{Geometry of the Resolution}  \label{ss:geom2}

The quotient singularity $\C^2/G$ is resolved by an ALE space which we
denote $X$. We want
to count the number of deformations of the tangent bundle of this
space. This problem has been solved by Kronheimer and Nakajima
\cite{MR1075769}.

\begin{figure}
\begin{center}
\begin{tikzpicture}[x=2mm,y=2mm,every node/.style={draw}]
  \path[shape=circle,inner sep=1pt]
     (0,0) node(a0) {$w_2$}
     (27.32,0) node(a1) {$w_1$}
     (13.66,23.66) node(a2) {$w_0$}
     (8.66,5) node(a3) {$v_2$}
     (18.66,5) node(a4) {$v_1$}
     (13.66,13.66) node(a5) {$v_0$};
  \draw[<->] (a0) -- (a3);
  \draw[<->] (a3) -- (a4);
  \draw[<->] (a4) -- (a5);
  \draw[<->] (a2) -- (a5); 
  \draw[<->] (a5) -- (a3);
  \draw[<->] (a4) -- (a1);
\end{tikzpicture}
\end{center}
\caption{The Kronheimer--Nakajima quiver for $A_2$.} \label{fig:KN}
\end{figure}
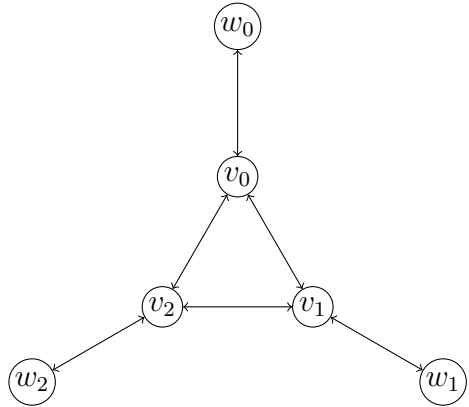

The moduli space of (Yang--Mills connections on) the bundle is given
by the moduli space of a quiver representation. To construct this
quiver one begins with the McKay quiver for $G$ and then, for each
node, one adds a new node and a path each way between these two
nodes. For example, for $A_2$ we obtain figure~\ref{fig:KN}
where each double-headed arrow represents an arrow in both directions.

Let $V$ and $W$ be two finite-dimensional representations of $G$. Let
$R_i$ be the irreducible representations of $G$. The decompositions of 
$V$ and $W$ are then 
\begin{equation}
  V = \bigoplus_i R_i^{\oplus v_i},\quad
  W = \bigoplus_i R_i^{\oplus w_i}.
\end{equation}

The dimension vector for the quiver representation is given by $v_i$
and $w_i$. The $v_i$'s are the dimensions for the McKay quiver and the
$w_i$'s correspond to the added nodes as in figure~\ref{fig:KN}.

Let $Q$ be the two-dimensional representation given by the embedding
$G\subset\Sl(2,\C)$. Kronheimer and Nakajima then construct the
desired bundle $E$ as the cohomology of a complex
\begin{equation}
\xymatrix@1{
  (V\otimes\cR)^G\ar[r]&(Q\otimes V\otimes \cR)^G\oplus
   (W\otimes\cR)^G\ar[r]&(\Lambda^2Q\otimes V\otimes\cR)^G,} \label{eq:KN1}
\end{equation}
where $\cR$ is a naturally defined bundle on $X$ which breaks up as a
sum of so-called tautological bundles $\cR_i$. We refer to
\cite{MR1075769} for details of the construction. The data giving the
maps in (\ref{eq:KN1}) come from the data of the representation of the
quiver subject to constraints coming from the ADHM equation.

The representation $W$ yields the asymptotic behaviour of the bundle
$E\to X$. Since we are looking at the tangent bundle this is simply
the representation $Q$. To determine the representation $V$ we need to
compute the Chern classes of the tangent bundle. Obviously $c_1(E)=0$.

To compute $c_2$ of the tangent bundle we use the Gauss-Bonnet formula
for a manifold with boundary given by Chern \cite{MR0014760}:
\begin{equation}
\chi(M) = \int_M c_{\dim M} + \int_{\partial M}\Pi,
\end{equation}
where $\Pi$ is the Chern--Simons form for the tangent bundle. For
$M$ given by a 4-dimensional ball we immediately obtain
\begin{equation}
\int_{S^3}\Pi = 1.
\end{equation}
Thus, if $X$ is an ALE space with asymptotic holonomy $G$ we have
\begin{equation}
\chi(X) = \int_X c_2(T_X) + \frac1{|G|}. \label{eq:c2}
\end{equation}
This Euler characteristic is given by the Euler characteristic of the
exceptional set and is therefore $n+1$ for $A_n$, $D_n$ or $E_n$.

Following \cite{MR1075769} we introduce four $(n+1)$-dimensional
vectors indexed by $i=0,\ldots,n$ where the 0 index is associated to
the trivial representation. $\mathbf{v}$ is defined with components
$v_i$, $\mathbf{w}$ with components $w_i$ and $\mathbf{d}$ (denoted
$\mathbf{n}$ in \cite{MR1075769}) has components given by the dimensions
of the irreducible representations of $G$. Finally $\mathbf{u}$ is
defined by the components $u_i$ given by
\begin{equation}
\begin{split}
  c_1(E) &= \sum_{i\neq0} u_i c_1(\cR_i)\\
  u_0 &= \mathbf{d}.\mathbf{w} - \sum_{i\neq0} d_iu_i.
\end{split}
\end{equation}
In the case of the tangent bundle $W=Q$ and thus $\mathbf{u}$ has
components $(2,0,0,\ldots)$. There is a relationship
\begin{equation}
  C\mathbf{v} = \mathbf{w}-\mathbf{u},  \label{eq:Cv}
\end{equation}
where $C$ is the Cartan matrix for the extended Dynkin diagram of
$G$. This determines $\mathbf{v}$ up to the kernel of $C$, which is
given by $\mathbf{d}$. The relation
\begin{equation}
  -\int_X\ch_2(E) = -\sum u_i\int_X\ch_2(\cR_i) + \frac{\dim V}{|G|},
\end{equation}
fixes this ambiguity. In our case, using (\ref{eq:c2}), this equation becomes 
\begin{equation}
  \dim V = (n+1)|G| - 1.\label{eq:dimV}
\end{equation}
Using $|G|=\sum d_i^2$, (\ref{eq:Cv}) and (\ref{eq:dimV}) fixes
\begin{equation}
  \mathbf{v} = (n+1)\mathbf{d} - (1,0,0,\ldots).
\end{equation}

The dimension of the moduli space of a quiver is easy to
determine. The dimension vectors $v_i$ and $w_i$ determine the sizes
of the matrices associated to each arrow. From the total number of
matrix entries one then subtracts the number of relations and finally
one subtracts the dimensions of the $\Gl(-,\C)$ actions on each node.

Since $X$ is a noncompact space we need to worry about boundary
conditions on the bundle. A natural choice is to consider a ``framed
bundle'' by fixing the asymptotic behaviour. The Kronheimer--Nakajima
quiver makes the framing picture very clean. The asymptotic form of
the bundle is determined purely by $W$. The $\Gl$ transformations
on the nodes associated with $W$ rotate this framing. In
order to provide a fixed framing for the bundle we therefore {\em
  ignore\/} the $\Gl$ action on these nodes. So, when computing the
dimension of the moduli space, we only consider the $\Gl$ action on
the nodes associated to $V$.

At the end of section 9 of \cite{MR1075769} the quaternionic
dimension for framed bundles is computed as
\begin{equation}
\dim\cM = \ff12\mathbf{v}.(\mathbf{w}+\mathbf{u}).
\end{equation}
Plugging in the values for the tangent bundle obtained above this
yields
\begin{equation}
\dim\cM = 2n+1.
\end{equation}

Comparing to theorem~\ref{th:orb2}, this proves the (0,2)-McKay
correspondence for dimension two.

\subsection{Compact Examples}

A K3 surface can be constructed as the resolution of an orbifold $S/G$
where $S$ is a 4-torus or another K3 surface. We should check that the
(0,2)-McKay correspondence works in these compact examples. The na\"\i
ve statement would be that the number of deformations of the tangent
bundle on K3 should equal the number of $G$-invariant deformations on
$S$ plus a contribution of $2n+1$ from each ADE-singularity.

This is not quite true --- there is a correction as observed in
\cite{AP:elusive}. We may compute the dimension of the moduli space of
the tangent bundle of a K3 surface easily by index theory:
\begin{equation}
\begin{split}
\sum_j (-1)^j\dim H^j(\End(T)) &= 2 - 2\dim\cM\\
  &= \int_X \ch(T)\wedge\ch(T)\wedge\td(T)\\
  &= -88.
\end{split}
\end{equation}
It follows that $\dim\cM=45$.

Let $S/G$ have global holonomy $H$ (i.e., $H\cong G$ if $S$ is a
4-torus and $H\cong\SU(2)$ if $S$ is a K3 surface) and let $Z(H)$ be
the centralizer of $H\subset \SU(2)$. A heterotic string compactified
on $S/G$ will have gauge group $E_8\times E_7\times Z(H)$. The process
of blowing up the singularities will Higgs away $Z(H)$ which will eat
up $\dim Z(H)$ scalars. The correct prediction is therefore
\begin{equation}
  45 = \dim \cM(T_S)^G + \sum_j(2n_j+1) - \dim Z(H),  \label{eq:K3}
\end{equation}
where the sum is over the $A_n$, $D_n$, $E_n$ singularities of $S/G$.

Consider, for example, the case where $S$ is a 4-torus and
$G\cong\Z_3$. Let $S$ have holomorphic coordinates $(z^1,z^2)$ and let
$G$ be generated by the action $g:(z^1,z^2)\mapsto(\omega z^1,\omega^2
z^2)$. $H^1(\End(T_S))$ is spanned by differential forms
\begin{equation}
  d\bar z^{\bar\imath} \otimes dz^j \otimes \frac{\partial}{\partial z^k}.
\end{equation}
It is easy to check that two of these are invariant under $g$. This
amounts to $\cM(T_S)^G$ having one quaternionic dimension. The
centralizer of $G$ in $\SU(2)$ is $\GU(1)$ and the orbifold has nine
$A_2$ singularities. Equation (\ref{eq:K3}) is indeed satisfied.

\begin{table}
\[
\begin{array}{|c|c|c|c|c|}
\hline
G&\hbox{Singularities}&\dim \cM(T_S)^G&\sum_j(2n_j+1)&Z(H)\\
\hline
\Z_2&16A_1&0&48&\SU(2)\\
\Z_3&9A_2&1&45&\GU(1)\\
\Z_4&4A_3+6A_1&0&46&\GU(1)\\
\Z_6&A_5+4A_2+5A_1&0&46&\GU(1)\\
D_4&2D_4+3A_3+2A_1&0&45&\\
D_5&D_5+2A_2+3A_3+A_1&0&45&\\
E_6&E_6+D_4+4A_2+A_1&0&45&\\
\hline
\end{array}
\]
\caption{Quotients of the form $T^4/G$.} \label{tab:T4}
\end{table}
There are only seven possible orbifolds $T^4/G$ where $G$ fixes a
point in $T^4$. These are listed in table~\ref{tab:T4}. All satisfy
(\ref{eq:K3}).

It is interesting to ask how one would prove (\ref{eq:K3}) mathematically
without using the Higgsing argument from string theory. It
seems\footnote{The author thanks M.~Douglas for suggesting this.} that
the correction from $Z(H)$ arises from the bundle framing issue. For
the resolution of $T^4/\Z_2$, for example, there must be some $\SU(2)$
symmetry of the bundle that amounts to a simultaneous frame rotation
of each of the 16 local pictures of the tangent bundle on the ALE
space associated to $A_1$. Thus we over-count by 3 if we simply add up
the local contributions from each ALE space. The author does not know
how to make this argument rigorous.

One may also check that the prediction works for the cases where $S$
is a K3 surface. These are listed in \cite{MR1385511}. Let $c=\sum_j
n_j$ denote the contribution to the Picard number of the exceptional
set as in table 2 of \cite{MR1385511}. Let there be
$N_{\textrm{sing}}$ singularities in the K3 orbifold. Then
(\ref{eq:K3}) implies
\begin{equation}
  N_{\textrm{sing}} = 45 -2c -\dim \cM(T_S)^G.
\end{equation}
If $G$ is a large group it is reasonable to expect that $\dim
\cM(T_S)^G=0$. Indeed, one can check that $N_{\textrm{sing}}=45-2c$
for most of the examples that appear late in table 2 of
\cite{MR1385511}.


\section{Abelian Quotients in Dimension Three} \label{s:dim3}

The moduli space of vector bundles in dimension three can be
obstructed. Accordingly, there is also the possibility of worldsheet
instanton corrections to the superpotential and hence the dimension of
the moduli space. Therefore, there is no reason to expect the number
of first order deformations counted by massless states at the orbifold
should agree with the moduli count in the resolution. Nevertheless we
will find perfect agreement in many cases.

\subsection{Toric Geometry}   \label{s:toric}

We will restrict attention to abelian quotients so that we may use the
tools of toric geometry. As usual, we use the homogeneous coordinate
ring \cite{Cox:} as based on a short exact sequence:
\begin{equation}
\xymatrix@1@C=20mm{
0\ar[r]&\mathsf{M}\ar[r]^-{\cA}&\Z^{\oplus N}\ar[r]^-\Phi&\mathsf{D}\ar[r]&0
} \label{eq:toric}
\end{equation}
$\mathsf{M}$ is a lattice of rank $d$. $\cA$ is an $N\times d$
matrix. The rows of $\cA$ give the coordinates of a set of $N$ points
in the lattice $\mathsf{N}$ dual to $\mathsf{M}$. We will use the same
symbol $\cA$ to denote this point set. Each point is associated to a
homogeneous coordinate $x_i$, $i=1,\ldots,N$. We then have a
homogeneous coordinate ring $S=\C[x_1,\ldots,x_N]$.

$\mathsf{D}$ is an abelian group which induces an action on the
homogeneous coordinates as follows. Let $r=N-d$ denote the rank of
$\mathsf{D}$. Then $\mathsf{D}\otimes_\Z\C^*=(\C^*)^r$ acts on the
homogeneous coordinates with charges given by the matrix $\Phi$. If
$\mathsf{D}$ has a torsion part $G$, then, in addition to $(\C^*)^r$,
we have a finite group action of $G$ on the homogeneous coordinates
the charges of which are given by the kernel of the matrix $\cA^t$ acting
on $(\Q/\Z)^{\oplus N}$.

The fan $\Sigma$ is defined as a fan over a simplicial
complex with vertices $\cA$. This combinatorial information defines an
ideal $B\subset S$ as explained in \cite{Cox:}. The toric variety
$X_\Sigma$ is then given by the quotient
\begin{equation}
  X_\Sigma = \frac{\Spec S - V(B)}{(\C^*)^r\times G}.
\end{equation}

For example, we may build $\C^3/\Z_5$, where the generator of $\Z_5$
acts as $\exp\Bigl(2\pi i(\ff15,\ff15,\ff35)\Bigr)$ on $(x_1,x_2,x_3)$
as follows. We set $N=n=3$ and thus require a $3\times 3$ integral
matrix $\cA$ whose kernel from a right-action on $(\Q/\Z)^{\oplus 3}$
is generated by $(\ff15,\ff15,\ff35)$. Clearly
\begin{equation}
  \cA = \begin{pmatrix}5&-1&-3\\0&1&0\\0&0&1\end{pmatrix},
\end{equation}
suffices. In this case $\Sigma$ is comprised of a single cone over a
triangle and $B=S$.

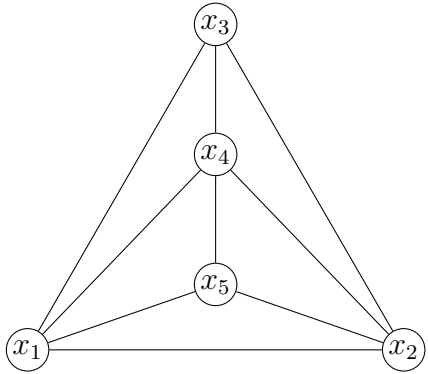
\begin{figure}
\begin{center}
\begin{tikzpicture}[x=1mm,y=1mm,every node/.style={draw}]
  \path[shape=circle,inner sep=1pt]
     (0,0) node(a0) {$x_1$}
     (50,0) node(a1) {$x_2$}
     (25,43.3) node(a2) {$x_3$}
     (25,26.0) node(a3) {$x_4$}
     (25,8.7) node(a4) {$x_5$};
  \draw (a0) -- (a1) -- (a2) -- (a0) -- (a4) -- (a3) -- (a2);
  \draw (a0) -- (a3) -- (a1) -- (a4);
\end{tikzpicture}
\end{center}
\caption{Toric Resolution of $\C^3/\Z_5$.}  \label{fig:Z5}
\end{figure}

In the case $X$ is a \CY\ variety, all the points $\cA$ lie in a
hyperplane in $\mathsf{N}$ . As is well-known, the orbifold
singularity $\C^3/G$ can be completely resolved crepantly by expanding the set
$\cA$ to contain all the points of $\mathsf{N}$ in its interior hull
and taking $\Sigma$ to be a fan over a simplicial complex including
all the new points of $\cA$. Let $\Delta$ denote this simplicial
complex. Each triangle in $\Delta$ has area $\ff12$,
$\mathsf{D}$ is torsion-free and $X_\Sigma$ is smooth.
For the above example of $\C^3/\Z_5$ we show the only choice of $\Delta$
in figure~\ref{fig:Z5}.

For a smooth toric variety, the homogeneous coordinate ring
$S=\C[x_1,\ldots,x_N]$ is multigraded by the free module $\mathsf{D}$:  
\begin{equation}
  S = \bigoplus_{\mathbf{d}\in\mathsf{D}} S_{\mathbf{d}}.
\end{equation}
Let $S(\mathbf{q})$ be the free $S$-module with grades shifted so that
$S(\mathbf{q})_{\mathbf{d}}=S_{\mathbf{q}+\mathbf{d}}$ as usual. Line
bundles or invertible sheaves $\O(\mathbf{q})$ on $X_\Sigma$ are then
associated to modules $S(\mathbf{q})$ for any
$\mathbf{q}\in\mathsf{D}$.

Let $\mathbf{q}_i$ denote the multi-degree of $x_i$. It was shown in
\cite{BatCox:toric} that the tangent sheaf $\cT$ of $X_\Sigma$ is given
by
\begin{equation}
\xymatrix@1@C=15mm{
0\ar[r]&\O^{\oplus r}\ar[r]^-E&\bigoplus_{i=1}^N \O(\mathbf{q}_i)
\ar[r]&\cT\ar[r]&0,
} \label{eq:torictan}
\end{equation}
where $E$ is an $N\times r$ matrix whose $(i,j)$-th entry is
$\Phi_{ji}x_i$.

We want to analyze the moduli space of the tangent sheaf. The first
order deformations are given by the vector space $\Ext^1(\cT,\cT)$ but
there may be obstructions. We refer to \cite{Hartshorne:def} for a
thorough review of these facts. 

In analogy with the Kronheimer--Nakajima construction we need to
identify the asymptotic form of the tangent bundle away from the
exceptional set. For the time being let us assume that $\C^3/G$ has an
isolated singularity at the origin.  Since $(x_1,x_2,x_3)$ were
coordinates on the orbifold prior to resolution it is natural to
define the sheaf
\begin{equation}
 \cW = \O(\mathbf{q}_1)\oplus\O(\mathbf{q}_2)\oplus\O(\mathbf{q}_3),
\end{equation}
which asymptotically resembles the tangent sheaf away from the origin.

The inclusion map $\cW\to \bigoplus_{i=1}^N \O(\mathbf{q}_i)$ induces
a map $f:\cW\to\cT$. The cokernel of this
map is isomorphic to the cokernel of the map
\begin{equation}
\xymatrix@1{
\O^{\oplus r}\ar[r]^-{E'}&\bigoplus_{\alpha=4}^N \O(\mathbf{q}_{\alpha}),
} \label{eq:tan1}
\end{equation}
where the matrix $E'$ is the matrix $E$ with the first three rows
removed. From now on, the index $\alpha$ will always be in the set
$4,\ldots,N$, and $N=r+3$.

Let $\Phi'$ be the matrix $\Phi$ with the first three columns
removed. One can show the square matrix $\Phi'$ must be invertible as
follows. Suppose it were not. Then there would be set of row
operations which would render a row all zero. The same row operations
applied to $\Phi$ would imply that there is a linear relation between
the coordinates of $x_1$, $x_2$ and $x_3$. This cannot be true as they
are the vertices of a triangle.  Since $\Phi'$ is invertible, by a
change of basis we may replace $E'$ by the diagonal matrix
$\diag(x_4,x_5,\ldots,x_{r+3})$.

Let $D_i$ be the toric divisor given by $x_i=0$, $i=1,\ldots,N$. The
exceptional divisor then has $r$ irreducible components given by
$D_\alpha$, $\alpha=4,\ldots,N$. The short exact sequences
\begin{equation}
\xymatrix@1{
  0\ar[r]&\O(-\mathbf{q}_\alpha)\ar[r]^-{x_\alpha}&\O\ar[r]&\O_{D_\alpha}
        \ar[r]&0,
}
\end{equation}
show that the cokernel of the map in (\ref{eq:tan1}) is isomorphic to
$\bigoplus_\alpha\O_{D_\alpha}(\mathbf{q}_\alpha)$. This proves
\begin{theorem} \label{th:tan}
  The tangent sheaf $\cT$ of the toric resolution of $\C^3/G$ is given
  by an extension
\begin{equation}
\xymatrix@1{
  0\ar[r]&\bigoplus_{i=1}^3\O(\mathbf{q}_i)\ar[r]&
        \cT\ar[r]&\bigoplus_{\alpha=4}^{N}
             \O_{D_\alpha}(\mathbf{q}_\alpha)\ar[r]&0.
} \label{eq:T-ext}
\end{equation}
\end{theorem}

\subsection{Quivers} \label{s:quivers}

From theorem~\ref{th:tan} the tangent bundle is a deformation of the
direct sum
\begin{equation}
  \bigoplus_{i=1}^3\O(\mathbf{q}_i)\oplus\bigoplus_{\alpha=4}^{N}
             \O_{D_\alpha}(\mathbf{q}_\alpha).  \label{eq:bigsum}
\end{equation}
A quiver is associated with this sum in a standard way
\cite{Ben:quiv}. That is, each summand $\Scr{V}_i$ is associated with
a node for $i=1,\ldots,N$. Then $\dim\Ext^1(\Scr{V}_i,\Scr{V}_j)$
arrows are drawn from node $i$ to node $j$.

The tangent bundle then corresponds to a representation of this
quiver. According to the precise form of the extension
(\ref{eq:T-ext}), we associate nonzero values to certain arrows
representing $\Ext^1(\O_{D_\alpha}(\mathbf{q}_\alpha),\O(\mathbf{q}_j))$. We now
explicitly construct this quiver.

\begin{theorem} \label{th:Ext} The groups
  $\Ext^1(\O_{D_\alpha}(\mathbf{q}_\alpha),\O_{D_\beta}(\mathbf{q}_\beta))$ and
  $\Ext^1(\O_{D_\alpha}(\mathbf{q}_\alpha),\O(\mathbf{q}_j))$ are determined by the
  normal bundle of the curve $C_{\alpha\beta}$ or $C_{\alpha j}$
  respectively, where $C_{ij}=D_i\cap D_j$. If $N_i$ is the normal
  bundle of the embedding $C_{ij}\subset D_i$ then
\begin{equation}
\begin{split}
  \dim\Ext^1(\O_{D_\alpha}(\mathbf{q}_\alpha),\O_{D_\beta}(\mathbf{q}_\beta)) &= h^0(C_{\alpha\beta},N_\alpha)\\
  \dim \Ext^1(\O_{D_\alpha}(\mathbf{q}_\alpha),\O(\mathbf{q}_j)) &= 1 + h^0(C_{\alpha
    j},N_\alpha).\\
  \dim\Ext^1(\O(\mathbf{q}_j),\O_{D_\alpha}(\mathbf{q}_\alpha)) &= 0
\end{split} \label{eq:Exts
}
\end{equation}
\end{theorem}

The proof is
facilitated by the language of the derived category. In what follows,
an underline represents position zero in a complex and we freely
identify a sheaf with a complex of sheaves with an entry only in
position zero. A shift of a complex $n$ places left is denoted $[n]$.

First note that $\O_{D_\alpha}(\mathbf{q}_\alpha)$ is $\xymatrix@1{\O\ar[r]^-{x_\alpha}&\poso{\O(\mathbf{q}_\alpha)}}$.
So $\sHom(\O_{D_\alpha}(\mathbf{q}_\alpha),\O_{D_\beta}(\mathbf{q}_\beta))$ is
\begin{equation}
\xymatrix@C=12mm{
  \O(-\mathbf{q}_\alpha)\ar[r]^-{\left(\begin{smallmatrix}-x_\alpha\\x_\beta\end{smallmatrix}\right)}&
    \poso{{\begin{matrix}\O\\\oplus\\\O(\mathbf{q}_\beta-\mathbf{q}_\alpha)\end{matrix}}}
       \ar[r]^-{\left(\begin{smallmatrix}
       x_\beta&x_\alpha\end{smallmatrix}\right)}&\O(\mathbf{q}_\beta).} \label{eq:sHom}
\end{equation}
But this is exactly $\O_{C_{\alpha\beta}}(\mathbf{q}_\beta)[-1]$, where $C_{\alpha\beta}=D_\alpha\cap
D_\beta$. So
\begin{equation}
\begin{split}
  \Ext^1(\O_{D_\alpha}(\mathbf{q}_\alpha),\O_{D_\beta}(\mathbf{q}_\beta)) &= H^0(C_{\alpha\beta},\O(\mathbf{q}_\beta))\\
     &= H^0(C_{\alpha\beta},N_\alpha),
\end{split}
\end{equation}
where $N_\alpha$ is the normal bundle of $C_{\alpha\beta}\subset
D_\alpha$. It should be noted that if $C_{\alpha\beta}$ is the empty
set then (\ref{eq:sHom}) is an exact complex and thus the above Ext
group vanishes.

Similarly $\sHom(\O_{D_\alpha}(\mathbf{q}_\alpha),\O(\mathbf{q}_j))$ is
\begin{equation}
\xymatrix@C=12mm{
  \poso{\O(\mathbf{q}_j-\mathbf{q}_\alpha)}\ar[r]^-{x_\alpha}&\O(\mathbf{q}_j)}
\end{equation}
which equals $\O_{D_\alpha}(\mathbf{q}_j)[-1]$. So
\begin{equation}
\Ext^1(\O_{D_\alpha}(\mathbf{q}_\alpha),\O(\mathbf{q}_j)) = H^0(\O_{D_\alpha}(\mathbf{q}_j)).
\end{equation}

But we have an exact sequence
\begin{equation}
\xymatrix@1{
0\ar[r]&\O_{D_\alpha}\ar[r]^-{x_j}&\O_{D_\alpha}(\mathbf{q}_j)\ar[r]&\O_{C_{\alpha
    j}}(\mathbf{q}_j)\ar[r]&0} \label{eq:aj1}
\end{equation}
which, since $H^1(\O_{D_\alpha})=0$
(as $D_\alpha$ is rational), implies
\begin{equation}
\dim \Ext^1(\O_{D_\alpha}(\mathbf{q}_\alpha),\O(\mathbf{q}_j)) = 1 + \dim H^0(C_{\alpha j},N_\alpha).
\label{eq:aj2}
\end{equation}

Next, $\Ext^1(\O(\mathbf{q}_j),\O_{D_\alpha}(\mathbf{q}_\alpha)) =
H^1(\O_{D_\alpha}(\mathbf{q}_\alpha-\mathbf{q}_j))$. We have a short exact sequence
\begin{equation}
\xymatrix@1{
0\ar[r]&\O_{D_\alpha}(\mathbf{q}_\alpha-\mathbf{q}_j)\ar[r]^-{x_j}&\O_{D_\alpha}(\mathbf{q}_\alpha)\ar[r]&
   \O_{C_{\alpha j}}(\mathbf{q}_\alpha)\ar[r]&0.}
\end{equation}
Noting that $\O_{D_\alpha}(\mathbf{q}_\alpha)$ is the canonical sheaf $\cK_{D_\alpha}$ of $D_\alpha$,
we have by Serre duality that
$H^0(\O_{D_\alpha}(\mathbf{q}_\alpha))=H^2(\O_{D_\alpha})=0$ and 
$H^1(\O_{D_\alpha}(\mathbf{q}_\alpha))=H^1(\O_{D_\alpha})=0$. 
The long exact sequence then gives
\begin{equation}
  \Ext^1(\O(\mathbf{q}_j),\O_{D_\alpha}(\mathbf{q}_\alpha)) = H^0(\O_{C_{\alpha
      j}}(N_j)). \label{eq:qjEa}
\end{equation}

To compute the proof of theorem \ref{th:Ext} we need to show that
$H^0(\O_{C_{\alpha j}}(N_j))=0$. The dimensions of the groups
$H^0(\O_{C_{ij}}(N_j))$ can be read directly from the toric data,
i.e., the triangulation of the pointset $\cA$ as in figure
\ref{fig:Z5}. The following result is standard in toric geometry. See,
for example, \cite{Fulton:}.

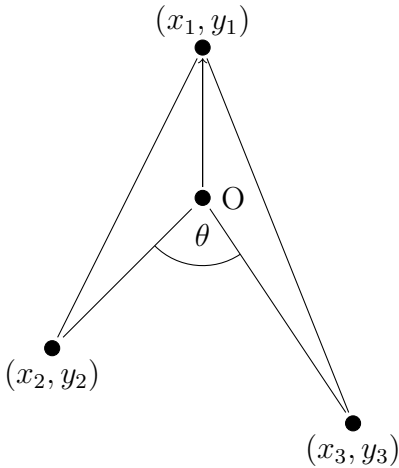
\begin{figure}
\begin{center}
\begin{tikzpicture}[x=1mm,y=1mm]
  \filldraw  (20,30) circle (1) node (a0) {} node [right=1mm] {O};
  \filldraw  (0,10) circle (1) node (a2) {} node [below] {$(x_2,y_2)$};
  \filldraw  (40,0) circle (1) node (a3) {} node [below] {$(x_3,y_3)$};
  \filldraw  (20,50) circle (1) node (a1) {} node [above] {$(x_1,y_1)$};
  \draw[<-] (a0) -- (a1) -- (a2) -- (a0) -- (a3) -- (a1) -- (a0);
  \draw (25,22.5) arc (303.7:225:9);
  \draw (20,25) node {$\theta$};
\end{tikzpicture}
\caption{Toric diagram for $C_{ij}=D_i\cap D_j$.} \label{fig:C}
\end{center}
\end{figure}

Let $D_i$ and $D_j$ correspond to two points $a_i$ and $a_j$ in
$\cA$. If $D_i$ and $D_j$ intersect at all it is along a $\P^1$
corresponding to a line joining $a_i$ and $a_j$ in the
triangulation. Choose integral coordinates in the plane containing
$\cA\subset\mathsf{N}$ such that $a_i$ becomes the origin and $a_j$
has coordinates $(x_1,y_1)$. Then
$C=D_i\cap D_j$ is described by the toric subfan in figure~\ref{fig:C}.
The normal bundle for $C$ is $\O(m)\oplus\O(-2-m)$ where
$m=x_2y_3-x_3y_2$ and, in particular, the normal bundle for $C\subset
D_i$ is $\O(m)$. 

If $\theta>\pi$ then $m<0$ and so $H^0(C,N_i)=0$. This is is the case
for (\ref{eq:qjEa}) since $D_j$ is a vertex of the triangle forming
the convex hull of the points in $\cA$. This completes the proof.
$\Box$

We should also note that figure~\ref{fig:C} gives a very quick method
of computing the dimension of the required $\Ext$ groups. These
numbers can read off the toric diagram describing the resolution. 
This first trivial observation is that $C_{ij}$ is only non-empty if
the nodes $a_i$ and $a_j$ are connected by a line.
{\em This means that the toric diagram itself becomes a quiver.}
The
rules are as follows.
\begin{itemize}
\item  For each pair of points $a_\alpha,a_\beta$ in the interior of the
  triangle joined by a line compute $m=x_2y_3-x_3y_2$ as in
  figure~\ref{fig:C}. If $m=-1$ there is no contribution, otherwise we
  have an arrow of multiplicity $m+1$ in the direction shown.
\item For each pair of points $a_\alpha$ in the interior and
  $\alpha_j$ as a vertex we do the following. If these points are
  joined by a line this becomes an arrow from $a_\alpha$ to $a_j$ of
  multiplicity $m+2$. If the points are not joined we have an arrow of
  multiplicity one from $a_\alpha$ to $a_j$.
\end{itemize}

The form of the extension corresponding to the tangent sheaf in
theorem~\ref{th:tan} tells us to give nonzero values to the arrows in
the quiver associated to the maps in
$\Ext^1(\O_{D_\alpha}(\mathbf{q}_\alpha),\O(\mathbf{q}_j))$ associated to multiplication
by $x_j$. From (\ref{eq:aj1}) this is the arrow associated with the
``1'' in the right-hand side of (\ref{eq:aj2}). All other arrows in
the quiver will be associated with a zero map for the quiver
representation yielding the tangent sheaf.

Now we have constructed the quiver, we need to compute the
relations. The obstruction theory of sheaves on a \CY\ threefold has
been explored in the context of string theory in the D-brane
literature and the same methods apply here. After all, both cases are
concerned with $N=1$ supersymmetry in four dimensions. We refer to
\cite{BDLR:Dq,AK:ainf} for further details. The key idea is that the
obstruction theory is governed by a superpotential that corresponds to
{\em loops\/} in the quiver. If there are no loops in the quiver then
the superpotential must be zero. It follows that the moduli space is
obstruction free. Most of the examples we consider have no loops. We
we consider the case of a loop in section \ref{s:superp}.

Finally, to compute the dimension of the moduli space we total the
number of degrees of freedom in the arrows of the quiver and subtract
the dimension of the $\Gl(-,\C)$ actions on each node. As in the case
of Kronheimer--Nakajima quiver, we do not include the $\C^*$-actions on
the three vertices of the triangle in order to fix a framing from the
bundle. Each of the interior points corresponds to a $\C^*$-action
that should be include. Thus the dimension of the moduli space is
equal to the number of arrows (with multiplicities) minus the number
of interior vertices.

Note that the number of interior points is equal to the number of
irreducible components of the exceptional set which, in turn, equals
the number of K\"ahler form deformations of the resolution. The total
number of singlets for the resolved isolated singularity equals this number
plus the number of deformations of the tangent sheaf. Thus we have a
statement for the geometric construction:

\begin{center}
 \shabox{The number of singlets is given by the number of arrows in the quiver
(including multiplicities) constructed from the toric diagram of the
resolution as above.}
\end{center}

This count corresponds to the non-linear $\sigma$-model computation
and, as such, can be corrected by worldsheet instanton effects.

It should be noted that computations of moduli spaces of quivers
require knowledge of stability conditions. Indeed, for the
(0,2)-theory we should concern ourselves with $\mu$-stable holomorphic
bundles. However, it is clear that the tangent bundle on a \CY\
manifold that is not locally a product is stable and lives in the
interior of the space of stability conditions. This follows since the
Hermitian connection on the tangent bundle is
Hermitian-Yang-Mills. Furthermore, the bundle is not a direct sum and
so we are not on the boundary of the space of polystable
connections. Thus we may ignore issues of stability when dealing only
with first order deformations of the tangent bundle.


\section{Examples} \label{s:egs}

\subsection{$\C^3/\Z_{2m+1}$}

In complex dimension three, from (\ref{eq:qbar}) we demand $\bar q=-\ff12$.

We will first consider the quotient $\Z_{2m+1}$ generated by the
weights 
\begin{equation}
\left(\frac1{2m+1},\frac1{2m+1},\frac{2m-1}{2m+1}\right).
\end{equation}
For explicitness we begin with the case $m=2$ corresponding to to
$\C^3/\Z_5$ considered in figure~\ref{fig:Z5}. Here we have four
twisted sectors, twisted by $\ff15$, $\ff25$, $\ff35$ or $\ff45$
respectively. The $\ff15$ and $\ff25$-twisted sectors contribute only
$\bar q=-\ff12$ states while the $\ff35$ and $\ff45$-twisted sectors
contribute only $\bar q=\ff12$ states.

In the $\ff15$-sector we have, for the vacuum
\begin{equation}
\begin{split}
\nu_i&=(\ff15,\ff15,\ff35), \tilde\nu_i=(-\ff3{10},-\ff3{10},-\ff9{10})\\
q&=-\sum(\tilde\nu_i+\ff12)=0\\
\bar q&=\sum(\nu_i-\ff12)=-\ff12\\
E&=-\ff58+\ff12\sum\left(\nu_i(1-\nu_i)+\tilde\nu_i(1+\tilde\nu_i)\right)\\
 &= -\ff35
\end{split}
\end{equation}

\begin{table}
\renewcommand\arraystretch{1.3}
\begin{equation}
\begin{array}{|c|c|c|c|}
\hline
\ff15&E&q&\bar q\\
\hline
x_i&\ff15,\ff15,\ff35&0&0\\
\rho_i&\ff45,\ff45,\ff25&0&0\\
\gamma_i&\ff7{10},\ff7{10},\ff1{10}&-1&0\\
\bar\gamma_i&\ff3{10},\ff3{10},\ff9{10}&1&0\\
\hline
\end{array}\qquad
\begin{array}{|c|c|c|c|}
\hline
\ff25&E&q&\bar q\\
\hline
x_i&\ff25,\ff25,\ff15&0&0\\
\rho_i&\ff35,\ff35,\ff45&0&0\\
\gamma_i&\ff9{10},\ff9{10},\ff7{10}&-1&0\\
\bar\gamma_i&\ff1{10},\ff1{10},\ff3{10}&1&0\\
\hline
\end{array}
\end{equation}
\caption{Excited modes in $\C^3/\Z_5$ twisted sectors.}\label{tab:Z5e}
\end{table}

The eigenvalues for the excitations are shown in table~\ref{tab:Z5e}.
This gives 11 singlets:
\begin{itemize}
\item $x_1^3, x_1^2x_2, x_1x_2^2, x_2^3, x_3, \rho_3x_i, 
  \gamma_3\bar\gamma_ix_j$, for $i,j=1,2$.
\end{itemize}

In the $\ff25$-sector we have
\begin{equation}
\begin{split}
\nu_i&=(\ff25,\ff25,\ff15), \tilde\nu_i=(-\ff1{10},-\ff1{10},-\ff3{10})\\
q&=-1\\
\bar q&=-\ff12\\
E& =-\ff12.
\end{split}
\end{equation}

This yields 7 singlets:
\begin{itemize}
\item $x_i\bar\gamma_j, x_3\bar\gamma_3, x_3^2\bar\gamma_i$, for $i,j=1,2$.
\end{itemize}

This gives a total of 18 singlets. 

\def\arrj#1#2#3{
  \coordinate (m) at ($ (#1)!.5!(#2) $);
  \draw[->] (#1) -- (m);
  \draw (m) -- (#2);
  \node[inner sep=0pt,label=above:$\scriptstyle #3$] at (m) {}
}

\def\arrja#1#2#3{
  \coordinate (m) at ($ (#1)!.5!(#2) $);
  \draw[->] (#1) -- (m);
  \draw (m) -- (#2);
  \node[inner sep=0pt,label=right:$\scriptstyle #3$] at (m) {}
}

\def\arrjb#1#2#3{
  \coordinate (m) at ($ (#1)!.5!(#2) $);
  \draw[->] (#1) -- (m);
  \draw (m) -- (#2);
  \node[inner sep=0pt,label=left:$\scriptstyle #3$] at (m) {}
}

\def\arrjd#1#2#3{
  \coordinate (m) at ($ (#1)!.5!(#2) $);
  \draw[->] (#1) -- (m);
  \draw (m) -- (#2);
  \node[inner sep=0pt,label=below:$\scriptstyle #3$] at (m) {}
}

\begin{figure}[b]
\begin{center}
\begin{tikzpicture}[x=1mm,y=1mm]
  \path[shape=circle,inner sep=1pt,every node/.style={draw}]
     (0,0) node(a0) {1}
     (50,0) node(a1) {2}
     (25,43.3) node(a2) {3}
     (25,26.0) node(a3) {4}
     (25,8.7) node(a4) {5};
  \draw[dashed] (a0) -- (a1) -- (a2) -- (a0);
  \arrja{a3}{a2}{5};
  \arrj{a3}{a1}{2};
  \arrj{a3}{a0}{2};
  \arrj{a4}{a1}{3};
  \arrj{a4}{a0}{3};
  \arrja{a4}{a3}{2};
  \draw[->] (a4) .. controls (50,25) and (50,35) .. (a2);
  \node[inner sep=0pt,label=above:$\scriptstyle1$] at (40,37) {};
\end{tikzpicture}
\end{center}
\caption{Quiver for $\C^3/\Z_5$.}  \label{fig:Z5q}
\end{figure}
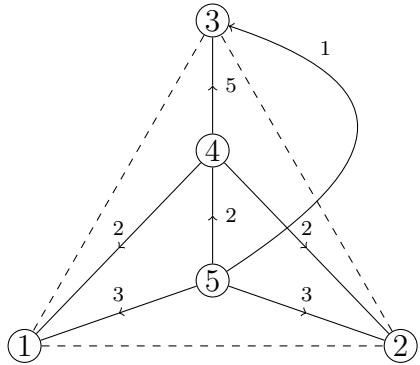

Applying the rules of section \ref{s:quivers}, the tangent sheaf
corresponds to the quiver shown in figure~\ref{fig:Z5q}. Adding up the
multiplicities on the arrows gives a total of 18. Thus the (0,2)-McKay
correspondence works for this example.

The general case $\C^3/\Z_{2m+1}$ is very similar. Assuming $m>1$ one can
compute the number of singlets in each sector:
\begin{itemize}
\item $\frac1{2m+1}$: $5+m(m+1)$
\item $\frac2{2m+1}$: 5
\\$\vdots$
\item $\frac{m-1}{2m+1}$: 5
\item $\frac{m}{2m+1}$: 7
\end{itemize}
giving a total of $m^2+6m+2$. This formula is also valid for $m=1$.

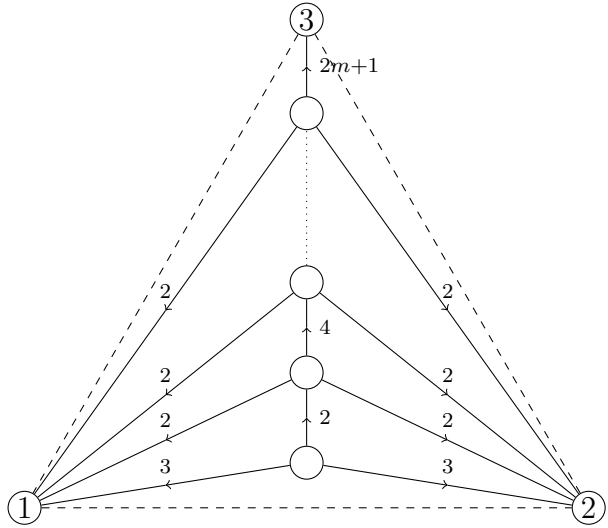
\begin{figure}
\begin{center}
\begin{tikzpicture}[x=1.5mm,y=1.5mm]
  \path[shape=circle,inner sep=1pt,every node/.style={draw}]
     (0,0) node(a0) {1}
     (50,0) node(a1) {2}
     (25,43.3) node(a2) {3}
     (25,35) node(a3) {$\phantom{1}$}
     (25,4) node(a4) {$\phantom{1}$}
     (25,12) node(a5) {$\phantom{1}$}
     (25,20) node(a6) {$\phantom{1}$};
  \draw[dashed] (a0) -- (a1) -- (a2) -- (a0);
  \arrja{a3}{a2}{2m+1};
  \arrj{a3}{a1}{2};
  \arrj{a3}{a0}{2};
  \arrj{a5}{a1}{2};
  \arrj{a5}{a0}{2};
  \arrj{a6}{a1}{2};
  \arrj{a6}{a0}{2};
  \arrj{a4}{a1}{3};
  \arrj{a4}{a0}{3};
  \arrja{a4}{a5}{2};
  \arrja{a5}{a6}{4};
  \draw[dotted] (a6) -- (a3);
\end{tikzpicture}
\end{center}
\caption{Quiver for $\C^3/\Z_{2m+1}$.}  \label{fig:Zmp1}
\end{figure}

The quiver is shown in figure~\ref{fig:Zmp1}. Note that for clarity we
have omitted the arrows of multiplicity one from interior points to
non-adjacent vertices. \textit{\textbf{We will always do this from now
    on.}}  Adding up all the multiplicities we again get $m^2+6m+2$.

\subsection{$\C^3/\Z_{11}$}  \label{ss:Z11}

The example of $\C^3/\Z_{2m+1}$ had no ambiguities in the
resolution. The simplest case of an isolated singularity with
ambiguities is $\C^3/\Z_{11}$ where the generator acts with weights
$(\ff1{11},\ff2{11},\ff8{11})$.

The conformal field theory computation yields a count of singlets
with $\bar q=-\ff12$ as follows:
\begin{itemize}
\item $\ff1{11}$: 14
\item $\ff2{11}$: 4
\item $\ff3{11}$: 5
\item $\ff6{11}$: 7
\item $\ff7{11}$: 9
\end{itemize}
for a total of 39.

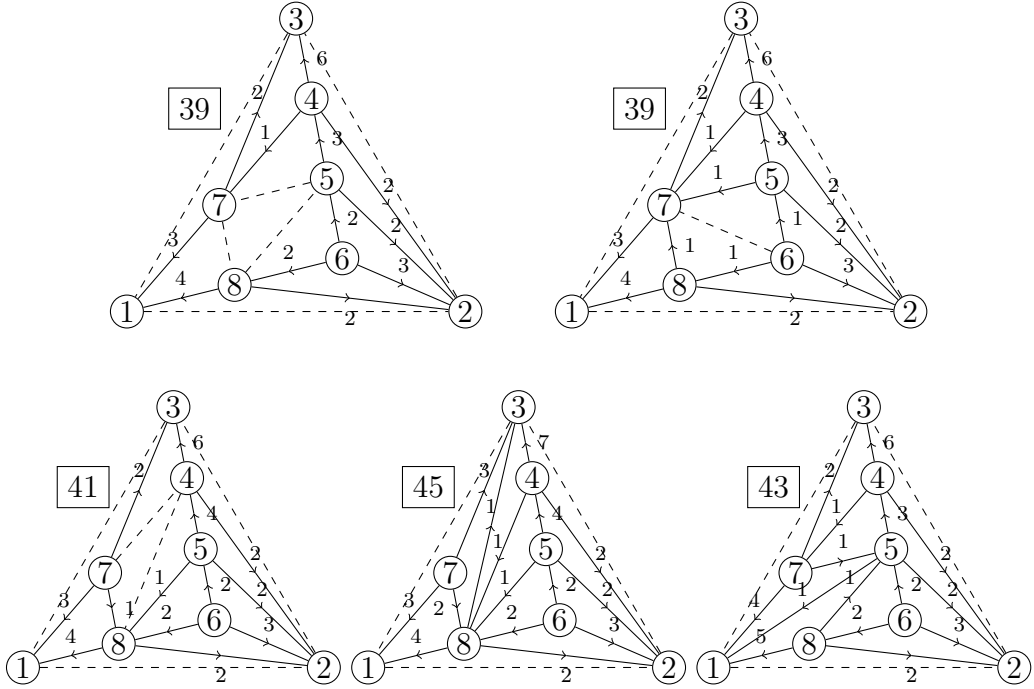
\begin{figure}
\begin{center}
\begin{tikzpicture}[x=0.9mm,y=0.9mm]
  \path[shape=circle,inner sep=1pt,every node/.style={draw}]
     (0,0) node(a0) {1}
     (50,0) node(a1) {2}
     (25,43.3) node(a2) {3}
     (27.27,31.49) node(a3) {4}
     (29.54,19.68) node(a4) {5}
     (31.81,7.87) node(a5) {6}
     (13.64,15.75) node(a6) {7}
     (15.91,3.94) node(a7) {8};
  \draw[dashed] (a0) -- (a1) -- (a2) -- (a0);
  \draw[dashed] (a4) -- (a6) -- (a7) -- (a4);
  \arrja{a3}{a2}{6};
  \arrj{a3}{a1}{2};
  \arrj{a4}{a1}{2};
  \arrj{a5}{a1}{3};
  \arrja{a5}{a4}{2};
  \arrja{a4}{a3}{3};
  \arrj{a3}{a6}{1};
  \arrj{a5}{a7}{2};
  \arrjd{a7}{a1}{2};
  \arrj{a7}{a0}{4};
  \arrj{a6}{a0}{3};
  \arrj{a6}{a2}{2};
  \draw[every node/.style={draw}] (10,30) node {39};
\end{tikzpicture}
\qquad
\begin{tikzpicture}[x=0.9mm,y=0.9mm]
  \path[shape=circle,inner sep=1pt,every node/.style={draw}]
     (0,0) node(a0) {1}
     (50,0) node(a1) {2}
     (25,43.3) node(a2) {3}
     (27.27,31.49) node(a3) {4}
     (29.54,19.68) node(a4) {5}
     (31.81,7.87) node(a5) {6}
     (13.64,15.75) node(a6) {7}
     (15.91,3.94) node(a7) {8};
  \draw[dashed] (a0) -- (a1) -- (a2) -- (a0);
  \draw[dashed] (a5) -- (a6);
  \arrj{a6}{a0}{3};
  \arrj{a7}{a0}{4};
  \arrj{a6}{a2}{2};
  \arrja{a3}{a2}{6};
  \arrj{a3}{a6}{1};
  \arrja{a4}{a3}{3};
  \arrja{a5}{a4}{1};
  \arrja{a7}{a6}{1};
  \arrj{a3}{a1}{2};
  \arrj{a4}{a1}{2};
  \arrj{a5}{a1}{3};
  \arrj{a4}{a6}{1};
  \arrj{a5}{a7}{1};
  \arrjd{a7}{a1}{2};
  \draw[every node/.style={draw}] (10,30) node {39};
\end{tikzpicture}\\[7mm]

\begin{tikzpicture}[x=0.8mm,y=0.8mm]
  \path[shape=circle,inner sep=1pt,every node/.style={draw}]
     (0,0) node(a0) {1}
     (50,0) node(a1) {2}
     (25,43.3) node(a2) {3}
     (27.27,31.49) node(a3) {4}
     (29.54,19.68) node(a4) {5}
     (31.81,7.87) node(a5) {6}
     (13.64,15.75) node(a6) {7}
     (15.91,3.94) node(a7) {8};
  \draw[dashed] (a0) -- (a1) -- (a2) -- (a0);
  \draw[dashed] (a6) -- (a3) -- (a7);
  \arrja{a3}{a2}{6};
  \arrj{a3}{a1}{2};
  \arrj{a4}{a1}{2};
  \arrj{a5}{a1}{3};
  \arrja{a5}{a4}{2};
  \arrja{a4}{a3}{4};
  \arrj{a5}{a7}{2};
  \arrjd{a7}{a1}{2};
  \arrj{a7}{a0}{4};
  \arrj{a6}{a0}{3};
  \arrj{a6}{a2}{2};
  \arrj{a4}{a7}{1};
  \arrja{a6}{a7}{1};
  \draw[every node/.style={draw}] (10,30) node {41};
\end{tikzpicture}
\begin{tikzpicture}[x=0.8mm,y=0.8mm]
  \path[shape=circle,inner sep=1pt,every node/.style={draw}]
     (0,0) node(a0) {1}
     (50,0) node(a1) {2}
     (25,43.3) node(a2) {3}
     (27.27,31.49) node(a3) {4}
     (29.54,19.68) node(a4) {5}
     (31.81,7.87) node(a5) {6}
     (13.64,15.75) node(a6) {7}
     (15.91,3.94) node(a7) {8};
  \draw[dashed] (a0) -- (a1) -- (a2) -- (a0);
  \arrja{a3}{a2}{7};
  \arrj{a3}{a1}{2};
  \arrj{a4}{a1}{2};
  \arrj{a5}{a1}{3};
  \arrja{a5}{a4}{2};
  \arrja{a4}{a3}{4};
  \arrj{a5}{a7}{2};
  \arrjd{a7}{a1}{2};
  \arrj{a7}{a0}{4};
  \arrj{a6}{a0}{3};
  \arrj{a6}{a2}{3};
  \arrj{a4}{a7}{1};
  \arrjb{a6}{a7}{2};
  \arrj{a3}{a7}{1};
  \arrj{a7}{a2}{1};
  \draw[every node/.style={draw}] (10,30) node {45};
\end{tikzpicture}
\begin{tikzpicture}[x=0.8mm,y=0.8mm]
  \path[shape=circle,inner sep=1pt,every node/.style={draw}]
     (0,0) node(a0) {1}
     (50,0) node(a1) {2}
     (25,43.3) node(a2) {3}
     (27.27,31.49) node(a3) {4}
     (29.54,19.68) node(a4) {5}
     (31.81,7.87) node(a5) {6}
     (13.64,15.75) node(a6) {7}
     (15.91,3.94) node(a7) {8};
  \draw[dashed] (a0) -- (a1) -- (a2) -- (a0);
  \arrja{a3}{a2}{6};
  \arrj{a3}{a1}{2};
  \arrj{a4}{a1}{2};
  \arrj{a5}{a1}{3};
  \arrja{a5}{a4}{2};
  \arrja{a4}{a3}{3};
  \arrj{a5}{a7}{2};
  \arrjd{a7}{a1}{2};
  \arrj{a7}{a0}{5};
  \arrj{a6}{a0}{4};
  \arrj{a6}{a2}{2};
  \arrj{a7}{a4}{1};
  \arrj{a4}{a0}{1};
  \arrj{a3}{a6}{1};
  \arrj{a6}{a4}{1};
  \draw[every node/.style={draw}] (10,30) node {43};
\end{tikzpicture}
\end{center}
\caption{Five quivers for the five resolutions of
  $\C^3/\Z_{11}$.}\label{fig:Z11}
\end{figure}

There are five possible resolutions of this orbifold given by five
different triangulations of the point set $\cA$. The quivers are shown
in figure~\ref{fig:Z11}. The number of singlets is shown in the square
box next to each diagram.

Only two of the five possible resolutions give the same number of
singlets as the conformal field theory. Thus, the (0,2)-McKay
correspondence is not true in a na\"\i ve sense. While the conformal
field theory is expected to give a precise count for the number of
singlets, the nonlinear sigma model may suffer from instanton
corrections which may decrease the number of singlets. Thus it should
come as no surprise that the geometrical computation may yield a
higher number than the conformal field theory. What is perhaps
surprising is that one always seems to need to work quite hard to find an
example where there really are instanton corrections
\cite{Beasley:2003fx,AP:elusive}.\footnote{It is conceivable that
  there are instanton corrections that kill moduli even when we do
  find an agreement in the count of singlets. That is, the
  superpotential obstructs deformations using massless modes away from
  the orbifold point. Then the same number of modes that were
  obstructed magically reappear at large radius when we ignore
  instanton corrections. We assume this is not the case but it would
  be nice to confirm this.}

It must therefore be that the last three diagrams in
figure~\ref{fig:Z11} contain rational curves which give corrections to
the superpotential along the lines described in
\cite{Dist:res,W:K3inst}.


\section{The $G$-Hilbert Scheme}  \label{s:GHilb}

For an orbifold $\C^3/G$, each triangulation of the point set $\cA$
leads to resolution of the singularity. There is, however, a
distinguished resolution called the $G$-Hilbert scheme \cite{MR1838978,IN:Hilb}.
This particular resolution has played a r\^ole in the McKay
correspondence \cite{BKM:MisM} but, in the context of (2,2)-models it
appears to have no distinguished r\^ole. Indeed, one of the
motivations of the analysis of topology change in string theory
\cite{AGM:II} was the fact that all possible crepant resolutions
should somehow be equal. We have just seen above, however, that such
egalitarianism does not extend to the (0,2) case.

Let $R=\C[x_1,x_2,x_3]$ and let $I\subset R$ be an ideal such that
$R/I$ is isomorphic to $\C^{\oplus|G|}$ as a vector space. Furthermore
let us demand that the action of $G$ on $R$ makes $R/I$ appear as the
regular representation of $G$. An obvious example of such an ideal is
\begin{equation}
(x_1-a_1,x_2-a_2,x_3-a_3)(x_1-b_1,x_2-b_2,x_3-b_3)\ldots,
\end{equation}
where $(a_1,a_2,a_3),(b_1,b_2,b_3),\ldots$ are the coordinates of a
free orbit of $G$ in $\C^3$. Less obvious examples of $I$ are
associated to orbits of $G$ with fixed points. The $G$-Hilbert scheme
parametrizes such ideals and it is shown in \cite{MR1783852} that it
provides a crepant resolution of the orbifold.

In the case where $G$ is abelian, the $G$-Hilbert scheme is toric and
so must correspond to some particular triangulation of the point set
$\cA$. This is determined as follows
\cite{MR1711344,Reid:McK1,Mohri:1998yd}.

Begin with the toric description of $\C^d$. $\mathsf{M}$ is then a
$d$-dimensional lattice. $\mathsf{M}$ may be viewed as the lattice of
characters for the $(\C^*)^d$-action on $\C^d$ \cite{Fulton:}. $G$ is
a subgroup of this $(\C^*)^d$-action. Let $\chi$ be a particular
character of $G$. The embedding $G\subset(\C^*)^d$ yields a subset
$\mathsf{M}_\chi\subset\mathsf{M}$ of characters corresponding to
$\chi$. Define $\mathsf{M}_\chi^+$ as the intersection of
$\mathsf{M}_\chi$ with the non-negative orthant.

For example, suppose $G$ is isomorphic to the cyclic group $\Z_n$ generated by the
action $\exp\left(\frac{2\pi i}n(a_1,a_2,\ldots,a_d)\right)$ on
$\C^d$, where the $a_i$ are integers. The characters of $\Z_n$
correspond to integers $j=0,\ldots,n-1$. We then define
\begin{equation}
\mathsf{M}_j^+ = \left\{\mathbf{m}\in(\Z_{\geq0})^d\:|\:
\mathbf{m}\cdot\mathbf{a}\equiv j\!\!\pmod{n}\right\}.
\end{equation}

Next define $\Sigma_\chi$ as the fan dual to the convex hull of
$\mathsf{M}_\chi^+$ and let $\Sigma_{G-\textrm{Hilb}}$ be the common
refinement of all the $\Sigma_\chi$'s as $\chi$ varies over all
characters of $G$. If $0$ is the trivial character, let $\mathsf{N}_0$
be dual to the lattice $\mathsf{M}_0$. The fan
$\Sigma_{G-\textrm{Hilb}}$ and the lattice $\mathsf{N}_0$ then provide
the toric data corresponding to the $G$-Hilbert scheme.

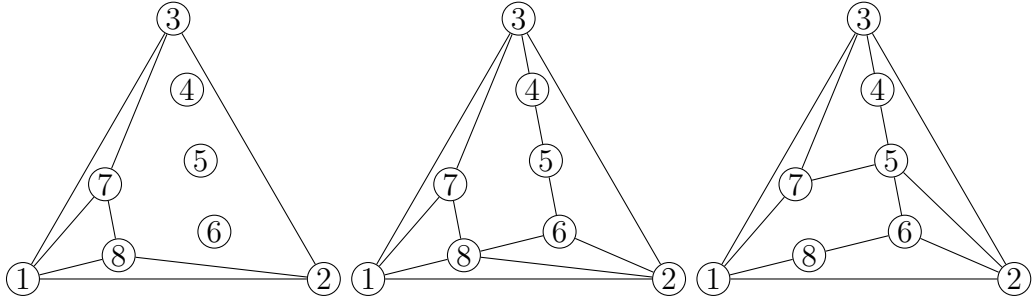
\begin{figure}
\begin{center}
\begin{tikzpicture}[x=0.8mm,y=0.8mm]
  \path[shape=circle,inner sep=1pt,every node/.style={draw}]
     (0,0) node(a0) {1}
     (50,0) node(a1) {2}
     (25,43.3) node(a2) {3}
     (27.27,31.49) node(a3) {4}
     (29.54,19.68) node(a4) {5}
     (31.81,7.87) node(a5) {6}
     (13.64,15.75) node(a6) {7}
     (15.91,3.94) node(a7) {8};
  \draw (a0) -- (a7) -- (a1) -- (a0) -- (a2) -- (a1);
  \draw (a0) -- (a6) -- (a7);
  \draw (a2) -- (a6);
\end{tikzpicture}
\begin{tikzpicture}[x=0.8mm,y=0.8mm]
  \path[shape=circle,inner sep=1pt,every node/.style={draw}]
     (0,0) node(a0) {1}
     (50,0) node(a1) {2}
     (25,43.3) node(a2) {3}
     (27.27,31.49) node(a3) {4}
     (29.54,19.68) node(a4) {5}
     (31.81,7.87) node(a5) {6}
     (13.64,15.75) node(a6) {7}
     (15.91,3.94) node(a7) {8};
  \draw (a0) -- (a7) -- (a5) -- (a4) -- (a3) -- (a2) -- (a1) -- (a0);
  \draw (a6) -- (a7) -- (a1) -- (a5);
  \draw (a0) -- (a2) -- (a6) -- (a0);
\end{tikzpicture}
\begin{tikzpicture}[x=0.8mm,y=0.8mm]
  \path[shape=circle,inner sep=1pt,every node/.style={draw}]
     (0,0) node(a0) {1}
     (50,0) node(a1) {2}
     (25,43.3) node(a2) {3}
     (27.27,31.49) node(a3) {4}
     (29.54,19.68) node(a4) {5}
     (31.81,7.87) node(a5) {6}
     (13.64,15.75) node(a6) {7}
     (15.91,3.94) node(a7) {8};
  \draw (a0) -- (a7) -- (a5) -- (a1) -- (a4) -- (a3) -- (a2) -- (a1)
           -- (a0);
  \draw (a5) -- (a4) -- (a6) -- (a2) -- (a0);
  \draw (a6) -- (a0);
\end{tikzpicture}
\end{center}
\caption{Slices of the fans $\Sigma_j$ for $j=1,3,7$ in the case of
  $\C^3/\Z_{11}$.}  \label{fig:Hfans}
\end{figure}

As an example let us consider the $\C^3/\Z_{11}$ case of section
\ref{ss:Z11}. We need to construct the fans
$\Sigma_0,\ldots,\Sigma_{10}$ for the 11 conjugacy classes. Each fan
is a fan over a triangulation of the point set $\cA$. In
figure~\ref{fig:Hfans} we show these triangulations in three cases. It
is a simple matter to compute these fans using a computer package
such as ``polymake''.

All said, when we combine these 11 fans together we obtain the
triangulation given by the first case in figure~\ref{fig:Z11}. This
was one of the two triangulations for which the (0,2)-McKay
correspondence did not require instanton corrections.

\subsection{Non-Isolated Singularities}  \label{ss:nonc}

Suppose the orbifold $\C^3/G$ is not isolated for $G\subset
\Sl(3,\C)$. Then there are fixed lines of singularities passing
through the origin. In both the conformal field and the geometric
picture the number of singlets is infinite.

In the conformal field theory description we will have bosonic
excitations of zero mass. Thus we may add arbitrary such excitations
to obtain massless singlets. In the geometric description there are
points in $\cA$ on the edges of the triangle forming the convex
hull. The $\Ext^1$ groups associated to arrows along such edges of the
triangle are infinite dimensional.

The lines of singularities emanating from the origin are locally of
the form $(\C^2/H)\times \C$, for some $H\subset\Sl(2,\C)$. We have
already proved the (0,2)-McKay correspondence for dimension two in
section \ref{s:dim2}. We should therefore be able to systematically
ignore the infinite number of states associated to two dimensions
leaving a finite number intrinsically associated with the
three-dimensional singularity at the origin. Let $N_0$
denote this finite number computed from the geometric picture.

For example, consider $\C^3/\Z_4$ where $\Z_4$ is generated by (the
exponential of) $(\ff14,\ff14,\ff12)$. The $\ff14$-twisted sector
contains 9 states with $\bar q=-\ff12$ while the $\ff34$-twisted
sector contains their $\bar q =\ff12$ partners. The $\ff12$-twisted
sector has an infinite number of states because we may have arbitrary
powers of $x_3$. Accordingly we ignore this sector. Thus we predict
$N_0=9$ if there are no instanton corrections.

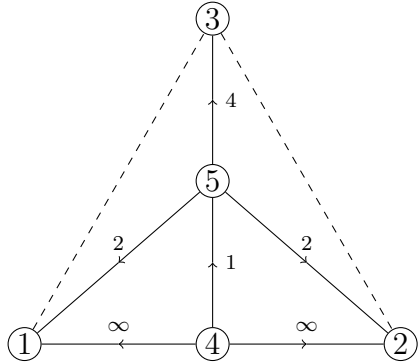
\begin{figure}[h]
\begin{center}
\begin{tikzpicture}[x=1mm,y=1mm]
  \path[shape=circle,inner sep=1pt,every node/.style={draw}]
     (0,0) node(a0) {1}
     (50,0) node(a1) {2}
     (25,43.3) node(a2) {3}
     (25,0) node(a3) {4}
     (25,21.7) node(a4) {5};
  \draw[dashed] (a0) -- (a2) -- (a1);
  \arrja{a3}{a4}{1};
  \arrja{a4}{a2}{4};
  \arrj{a4}{a0}{2};
  \arrj{a4}{a1}{2};
  \arrj{a3}{a0}{\infty};
  \arrj{a3}{a1}{\infty};
\end{tikzpicture}
\end{center}
\caption{Quiver for $\C^3/\Z_4$.}  \label{fig:Z4}
\end{figure}
The quiver for this case is shown in figure~\ref{fig:Z4}. Node 4 in
this figure represents a line of singularities of the form
$\C^2/\Z_2\times\C$. The arrows along the edges of the triangle
associated to this vertex are correspondingly infinite. We identify
this line of singularities as associated to the $\ff12$-twisted sector
in the conformal field theory. Adding up the finite arrow
multiplicities gives $N_0=9$ in agreement with the CFT count.

This allows us to extend our conjectured (0,2)-McKay correspondence to
cases where the singularities are non-isolated. 

\subsubsection{$\C^3/(\Z_m\times\Z_m)$}
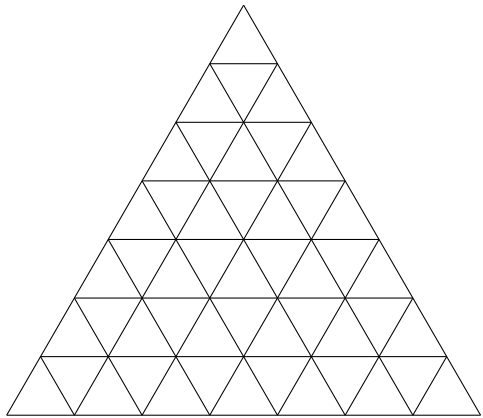
\begin{figure}[b!]
\begin{center}
\begin{tikzpicture}[x=9mm,y=9mm]
  \foreach \x in {0,...,6}
    \draw (0.5*\x,0.866*\x) -- (7-0.5*\x,0.866*\x);
  \foreach \x in {0,...,6}
    \draw (\x+1,0) -- (0.5*\x+0.5,0.866*\x+0.866);
  \foreach \x in {0,...,6}
    \draw (\x,0) -- (3.5 + 0.5*\x,-0.866*\x+7*0.866);
\end{tikzpicture}
\end{center}
\caption{$G$-Hilbert scheme for $\C^3/(\Z_7\times\Z_7)$.}  \label{fig:Z7x7}
\end{figure}

A particularly
symmetric case concerns the quotient $\C^3/(\Z_m\times\Z_m)$. Here the
$G$-Hilbert scheme is given torically by equilateral triangles
forming ``isometric graph paper''. We show the case for $m=7$ in
figure~\ref{fig:Z7x7}.

This singularity is not isolated; there are 3 lines of $\C^2/\Z_m$
emanating from the origin. These correspond to the three edges of the
triangle in figure~\ref{fig:Z7x7}. We will therefore ignore points on
these edges.

From the rules of section \ref{s:quivers} we see that the {\em only\/}
arrows in this diagram are the ones we have been omitting --- namely
the ones from interior points to non-adjacent vertices. Thus, the
number of deformations equals three times the number of strictly
interior points. That is, $\ff32(m-1)(m-2)$.

In the conformal field theory let us consider states in the sector
twisted by $g^ph^q$ where $g$ acts as $\exp 2\pi
i\left(\ff1m,\ff{m-1}m,0\right)$ and $h$ acts as $\exp 2\pi
i\left(0,\ff1m,\ff{m-1}m\right)$ on $\C^3$. If $p=0$, $q=0$ or $p=q$
then we have an infinite number of massless states. These correspond
to the lines of singularities. Otherwise, if $p<q$ then each sector
has 3 massless states with $\bar q=-\ff12$. Similarly, if $p>q$ then
we have 3 states with $\bar q=\ff12$. This gives again a total
of $\ff32(m-1)(m-2)$ states.

Actually, we have a stronger result:

\begin{prop}
Amongst all crepant resolutions of $\C^3/(\Z_m\times\Z_m)$, $N_0$
is minimized at $\ff32(m-1)(m-2)$ for the $G$-Hilbert scheme. All
other resolutions give a greater number. Thus, only the $G$-Hilbert
scheme is free from instanton corrections.
\end{prop}

To see this first note that the contribution to
$\Ext^1(\O_{D_\alpha}(\mathbf{q}_\alpha),\O(\mathbf{q}_j))$ in (\ref{eq:Exts }) is at
least $\ff32(m-1)(m-2)$ and that this lower bound is only achieved if
no interior point is connected to a vertex. Furthermore, from figure
\ref{fig:C}, the contribution to
$\Ext^1(\O_{D_\alpha}(\mathbf{q}_\alpha),\O_{D_\beta}(\mathbf{q}_\beta))$ is only zero
if all neighbouring pairs of triangles form strictly convex
quadrilaterals.  Starting from one corner of the big outer triangle
and working inwards, one can then see that these two conditions force
the triangulation to look like isometric graph paper.


\section{Superpotentials} \label{s:superp}

For three-dimensional examples above we have ignored the possibility
of relations. In the case of three dimensions such relations are
manifested in the form of a superpotential.

The ADHM relations on the Kronheimer--Nakajima quivers in the case of
two dimensions were very important to get the counting correct. The
source of such relations may be traced to the fact that every arrow is
paired with an arrow in the opposite direction because of Serre
duality. In fact, Serre duality itself is enough to derive the
superpotential (or its equivalent content) and thus the ADHM equations
but we do not include the details here.

In this paper we have thus far only been interested in counting the
first-order deformations. This count is affected by linear relations
or, equivalently, quadratic mass terms in the superpotential. Higher
order terms in the superpotential correspond to higher-order
obstructions to the first-order deformations. The ADHM relations from
the Kronheimer--Nakajima quivers would appear to be quadric but it is
important to remember that the tangent bundle corresponds to nonzero
values for maps on the arrows. Expanding about such nonzero values
makes the relations equivalent to linear relations and thus masses for
these deformations. Conversely, when we consider quivers associated
with three-dimensional cases, the arrows forming loops to form a
superpotential will be associated to arrows strictly in the interior
of the quiver. These maps are zero for the tangent bundle from
theorem~\ref{th:tan}. Thus we are expanding around zero and a cubic or
higher superpotential corresponds purely to obstructions.

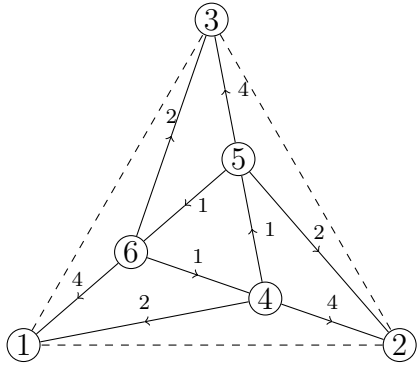
\begin{figure}
\begin{center}
\begin{tikzpicture}[x=1mm,y=1mm]
  \path[shape=circle,inner sep=1pt,every node/.style={draw}]
     (0,0) node(a0) {1}
     (50,0) node(a1) {2}
     (25,43.3) node(a2) {3}
     (32.14,6.19) node(a3) {4}
     (14.29,12.37) node(a4) {6}
     (28.57,24.74) node(a5) {5};
  \draw[dashed] (a0) -- (a2) -- (a1) -- (a0);
  \arrj{a4}{a3}{1};
  \arrja{a3}{a5}{1};
  \arrja{a5}{a4}{1};
  \arrj{a5}{a1}{2};
  \arrj{a3}{a1}{4};
  \arrj{a4}{a2}{2};
  \arrj{a4}{a0}{4};
  \arrj{a3}{a0}{2};
  \arrja{a5}{a2}{4};
\end{tikzpicture}
\end{center}
\caption{Quiver for $\C^3/\Z_7$.}  \label{fig:Z7}
\end{figure}

So far none of the quivers we have drawn have contained an oriented
cycle. Figure \ref{fig:Z7} depicts the case of $\C^3/\Z_7$. Here we do
indeed have an oriented cycle. This implies we have a possible
superpotential and thus that the moduli space can have
obstructions. Our goal in this section is to show that this cubic term
is nonzero. Note that we are computing the form of the superpotential
{\em geometrically\/} which is classical in terms of the non-linear
$\sigma$-model. So we are not considering worldsheet instanton
corrections.

The theory of computing superpotentials in $N=1$ theories in four
dimensions has been explored quite thoroughly in the context of
D-brane world-volumes, see for example, \cite{BDLR:Dq}. We may use
identical methods here. In particular, the superpotential encodes an
$A_\infty$-algebra as described in \cite{AK:ainf}. This $A_\infty$
structure becomes more apparent for terms in the superpotential higher
than cubic. However, in the case at hand we have a loop of length 3
and thus we are just dealing with a cubic term.

A loop around internal nodes $\alpha,\beta,\gamma$ corresponds to a
cubic term in the superpotential with coefficient given by the Yoneda
product
\begin{equation}
\begin{split}
  \Ext^1\left(\O_{D_{\alpha}}(\mathbf{q}_\alpha),\O_{D_{\beta}}(\mathbf{q}_\beta)\right)
  \times
  \Ext^1\left(\O_{D_{\beta}}(\mathbf{q}_\beta),\O_{D_{\gamma}}(\mathbf{q}_\gamma)\right)
  &\times\\
  \Ext^1\left(\O_{D_{\gamma}}(\mathbf{q}_\gamma),\O_{D_{\alpha}}(\mathbf{q}_\alpha)\right)
  &\to
  \Ext^3\left(\O_{D_{\alpha}}(\mathbf{q}_\alpha),\O_{D_{\alpha}}(\mathbf{q}_\alpha)\right)\\
  &\cong\C.
\end{split}
\end{equation}
The last equality here is obtained from Serre duality which we may use
as the sheaves are compactly supported. Using Serre duality again we may
rewrite this as
\begin{equation}
  \Ext^1\left(\O_{D_{\alpha}}(\mathbf{q}_\alpha),\O_{D_{\beta}}(\mathbf{q}_\beta)\right)
  \times
  \Ext^1\left(\O_{D_{\beta}}(\mathbf{q}_\beta),\O_{D_{\gamma}}(\mathbf{q}_\gamma)\right)
  \to
  \Ext^2\left(\O_{D_{\alpha}}(\mathbf{q}_\alpha),\O_{D_{\gamma}}(\mathbf{q}_\gamma)\right).
\end{equation}
This product is explicitly computed using the local cohomology
description of sheaf cohomology on toric varieties
\cite{EMS:ToricCoh}.

For this example, the matrix of charges in (\ref{eq:toric}) is
given by
\begin{equation}
\Phi=\begin{pmatrix}
0&0&1&1&-2&0\\
1&0&0&0&1&-2\\
0&1&0&-2&0&1
\end{pmatrix}
\end{equation}

The sheaf $\O_{D_{\alpha}}(\mathbf{q}_\alpha)$ is, in the derived category,
equivalent to the complex
$\xymatrix@1{\O\ar[r]^-{x_\alpha}&\O(\mathbf{q}_\alpha)}$.  The required Ext's
can therefore be computed via a spectral sequence in terms of sheaf
cohomology of line bundles. To be precise,
\begin{equation}
\begin{split}
  \Ext^n(\O_{D_\alpha}(\mathbf{q}_\alpha),\O_{D_\beta}(\mathbf{q}_\beta)) &=
    \bigoplus_{p+q=n}E^{p,q}_\infty,\\
  \textrm{with}\: E_1^{p,q} &= H^q(\cE^p),
\end{split}
\end{equation}
and $\cE^\bullet$ is the complex given in (\ref{eq:sHom}).  The
cohomology of line bundles is represented by Laurent monomials as
explained in \cite{meMP:singlets,Blumenhagen:2010pv,AP:elusive}. Such
computations were explained in gory detail in \cite{AP:elusive}.

For $\Ext^1\left(\O_{D_{4}}(\mathbf{q}_4),\O_{D_{5}}(\mathbf{q}_5)\right)$, the
only contribution comes from $\Ext^1\left(\O(\mathbf{q}_4),\O(\mathbf{q}_5)\right)\cong
H^1(\O(\mathbf{q}_5-\mathbf{q}_4))\cong H^1(\O(-3,1,2))$. This is represented by the
Laurent monomial
\begin{equation}
\frac{x_1}{x_3^2x_4}.
\end{equation}
Similarly $\Ext^1\left(\O_{D_{5}}(\mathbf{q}_5),\O_{D_{6}}(\mathbf{q}_6)\right)$ is
represented by $x_2/x_1^2x_5$. The Yoneda product of these two Ext
representatives is simply the product of these two monomials.

To compute $\Ext^2\left(\O_{D_{4}}(\mathbf{q}_4),\O_{D_{6}}(\mathbf{q}_6)\right)$ we
first note that $\Ext^2\left(\O(\mathbf{q}_4),\O(\mathbf{q}_6)\right)$ is 2-dimensional
and represented by monomials
\begin{equation}
\frac{x_2}{x_1x_3^2x_4x_5}\hbox{~and~}\frac{1}{x_1x_2x_3x_4^2x_5}.
\label{eq:E2x}
\end{equation}
However, $\Ext^2(\O,\O(\mathbf{q}_6))$ is one-dimensional and represented by
$1/x_1x_2x_3x_4x_5$. At the $E_1$ stage of the spectral sequence, the
second monomial in (\ref{eq:E2x}) is mapped to this by multiplication
by $x_4$ and so the second monomial in (\ref{eq:E2x}) is killed.

The result is that
$\Ext^2\left(\O_{D_{4}}(\mathbf{q}_4),\O_{D_{6}}(\mathbf{q}_6)\right)$ is
one-dimensional and is generated by the Yoneda product of the
generators of $\Ext^1\left(\O_{D_{4}}(\mathbf{q}_4),\O_{D_{5}}(\mathbf{q}_5)\right)$
and $\Ext^1\left(\O_{D_{5}}(\mathbf{q}_5),\O_{D_{6}}(\mathbf{q}_6)\right)$. Thus {\em
  the superpotential is a nonzero cubic corresponding to the loop
in figure~\ref{fig:Z7}.}
\begin{equation}
  W = XYZ.
\end{equation}
The derivatives of this superpotential imply that turning on one of
these three deformations obstructs the other two. 

Both the geometry and conformal field theory agree that there are 
24 singlets associated to $\C^3/\Z_7$. The appearance of a
superpotential does not change this count.


\section{Discussion}   \label{s:disc}

The agreement between the counting of states between the orbifold
conformal field theory and the deformations of the tangent bundle on
the resolved space clearly motivates the following:
\begin{conjecture}
  The counting of the number of states on a three-dimensional
  $G$-Hilbert scheme corresponding to $(0,2)$-deformations of an
  $N=(2,2)$ theory matches the conformal field theory orbifold count.
\end{conjecture}

We have proved this conjecture to be true above in an infinite number
of cases. Obviously it would be nice to check it in an even larger
class, such as all abelian orbifolds.

We have made no attempt in this paper to directly confront the
instanton computation. For isolated $\P^1$'s this amounts to computing
the splitting type of the bundle $E\to\P^1$ as $E$ is deformed away
from the tangent bundle \cite{Dist:res,W:K3inst}. This is not
particularly easy for the following reason. The tangent sheaf has a
nice presentation in terms of toric geometry in
(\ref{eq:torictan}). Deformations of the maps $E$ in this short exact
sequence will yield deformations of the tangent sheaf. Unfortunately
not all of the deformations can be understood so simply and it is
these extra deformations which appear to be volatile under flops
between different possible resolutions. Indeed the work of
\cite{Beasley:2003fx} implies that the we should expect all the
interesting instanton effects to be associated to these more obscure
deformations.

Another obvious unanswered question raised by the conjecture is ``Why
the $G$-Hilbert Scheme''? Is there some intrinsic reason why the
construction of the $G$-Hilbert scheme is guaranteed to reproduce the
orbifold computation? One thing that seems fairly likely is that, of
all the resolutions, the $G$-Hilbert scheme minimizes the number of
deformations. We proved this for $\C^3/(\Z_m\times\Z_m)$. More
generally the $G$-Hilbert scheme tries to get as close to isometric
graph paper as it can and thus minimizes the number of deformations.
For a more precise statement of this, see \cite{MR2075608}. Anyway, assuming
the $G$-Hilbert scheme minimizes the number of deformations, it is
therefore the ``most instanton free'' in some sense.

In this paper we have focused mainly on the instanton effects on
mass. We really have the whole superpotential to work with and we
showed in section \ref{s:superp} that there can be nontrivial
information here. It would be most interesting to compare conformal
field theory computations and geometrical computations of the
superpotential beyond the mass term.

\section*{Acknowledgments}

I thank M.~Douglas. I.~Melnikov and R.~Plesser for many useful and important
discussions. I would also like to thank the Simons Center for Geometry
and Physics for its hospitality and providing a stimulating
environment for much of this research. This work was partially
supported by NSF grant DMS--0905923.  Any opinions, findings, and
conclusions or recommendations expressed in this material are those of
the authors and do not necessarily reflect the views of the National
Science Foundation.


\end{document}